\documentclass[aps,nofootinbib,showkeys,noshowpacs,preprintnumbers,amsmath,amssymb]{revtex4-1}

\usepackage{amsmath,amsfonts,amssymb,amsthm,bm,bbm,bbold,braket,color,dsfont,fancybox,hyperref,
srcltx,subfigure,ucs,yfonts,cancel,xfrac,verbatim,xcolor}

\usepackage{float}

\usepackage{mathtools}
\usepackage[T1]{fontenc} 
\newcommand{\beq}{\begin{equation}}
\newcommand{\eeq}{\end{equation}}
\newcommand{\bea}{\begin{eqnarray}}
\newcommand{\eea}{\end{eqnarray}}
\newcommand{\beas}{\begin{eqnarray*}}
\newcommand{\eeas}{\end{eqnarray*}}
\newcommand{\bi}{\begin{itemize}}
\newcommand{\ei}{\end{itemize}}
\def\bes{\begin{subequations}}
\def\ees{\end{subequations}}

\usepackage{amsmath,amsfonts,amssymb,amsthm,bm,bbm,bbold,braket,color,dsfont,fancybox,hyperref,
srcltx,slashed,ucs,yfonts,cancel,xfrac,verbatim,xcolor}

\usepackage[inline,shortlabels]{enumitem}
\usepackage{nicefrac}
\usepackage{multirow}

\DeclareMathAlphabet{\mathpzc}{OT1}{pzc}{m}{it}
\definecolor{gold}{rgb}{1,0.8,0}
\definecolor{nara}{rgb}{1,0.4,0.1}
\definecolor{goldo}{rgb}{1,0.7,0}
\definecolor{greeno}{rgb}{0,0.8,0}
\def\bes{\begin{subequations}}
\def\ees{\end{subequations}}
\def\be{\begin{equation}}
\def\ee{\end{equation}}
\def\bea{\begin{eqnarray}}
\def\eea{\end{eqnarray}}
\def\ba{\begin{eqnarray}}
\def\ea{\end{eqnarray}}
\def\bear{\begin{array}}
\def\eear{\end{array}}

\newcommand{\bpm}{\begin{pmatrix}}
\newcommand{\epm}{\end{pmatrix}}
\newcommand{\BM}{\left(\begin{array}}		
\newcommand{\BMC}{\left[\begin{array}}		

\newcommand{\EM}{\end{array}\right)}		
\newcommand{\EMC}{\end{array}\right]}		

	% ...

\newcommand{\p}{\partial}

\newcommand{\com}[1]{}
\newcommand{\K}{\mathcal{K}}

\def\bes{\begin{subequations}}
\def\ees{\end{subequations}}
\def\be{\begin{equation}}
\def\ee{\end{equation}}
\def\bea{\begin{eqnarray}}
\def\eea{\end{eqnarray}}
\def\ba{\begin{eqnarray}}
\def\ea{\end{eqnarray}}
\def\bear{\begin{array}}
\def\eear{\end{array}}
\def\p1sl{\displaystyle{\not}p_1}
\def\p2sl{\displaystyle{\not}p_2}

\newcommand{\bG}{{\overline{\Gamma}}}

\newcommand{\Dst}{D^*}

\usepackage{mathtools}

\begin{document}


\title{Exploring CP-violation, via heavy neutrino oscillations, in rare $B$ meson decays at Belle II}

\author{Gorazd Cveti\v{c}$^{1}$}
\email{gorazd.cvetic@usm.cl}
\author{C.S. Kim$^{2}$}
\email{cskim@yonsei.ac.kr}
\author{Sebastian Mendizabal$^{1}$}
\email{sebastian.mendizabal@usm.cl}
\author{Jilberto Zamora-Sa\'a$^{3}$}
\email{jilberto.zamora@unab.cl}

 \affiliation{$^1$Department of Physics, Universidad T\'ecnica Federico Santa Mar\'ia, Valpara\'iso, Chile.}
  \affiliation{$^2$Department of Physics and IPAP, Yonsei University, Seoul 03722, Korea.}
 \affiliation{$^3$Departamento de Ciencias F\'isicas, Universidad Andres Bello,  Sazi\'e 2212, Piso 7,  Santiago, Chile.}
\begin{abstract}


In this article we study the rare B-meson decay via two on-shell almost-degenerate Majorana Heavy Neutrinos, into two charged leptons and two pseudoscalar mesons ($B^{\pm} \to D^0 \ell^{\pm}_1 \ell^{\pm}_2 \pi^{\mp}$). We consider the scenario where the heavy neutrino masses are $\sim 2$ GeV and the heavy-light mixing coefficients are $|B_{\ell N}|^2 \sim 10^{-5}$, and evaluate the possibility to measure the CP-asymmetry at Belle II. We present some realistic conditions under which the asymmetry could be detected.

\end{abstract}


\maketitle

\section{Introduction}
\label{s1}
The first indications of physics beyond the Standard Model (SM) come from: neutrino oscillations (NOs), baryonic asymmetry of the Universe (BAU) and dark matter (DM). During the last years NOs experiments have confirmed that active neutrinos ($\nu$) are very light massive particles ~\cite{Fukuda:1998mi,Eguchi:2002dm} and consequently the SM must be extended. The evidence of neutrino masses that arises from oscillations were first predicted in~\cite{Pontecorvo:1957qd} and later observed in~\cite{Fukuda:1998mi,Ahmad:2002jz,Lipari:2001ds,Rahman:2012cp,Dasgupta:2012dn}. 
These extremely light masses can be explained with the introduction of sterile neutrinos and via the seesaw mechanism~\cite{GellMann:1980vs,Sawada:1979dis,PhysRevLett.44.912}. The outcome gives us Majorana neutrinos with light eigenstates $m_{\nu}\lesssim$ 1 eV and heavy neutrino (HN) eigenstates. The masses of the HN particles are normally taken in the  $M_{N}\gg$ 1 TeV regime. However, there are other seesaw scenarios with lower masses for the HN,  $M_{N}\sim$ 1 TeV~\cite{Wyler:1982dd,Witten:1985xc,Mohapatra:1986bd,Mohapatra:1986bd,Malinsky:2005bi,Dev:2009aw,Dev:2012sg,Dev:2013oxa} and  $M_{N}\sim$ 1 GeV~\cite{Buchmuller:1991ce,Kohda2013,Asaka:2005an,Asaka:2005pn,delAguila:2007ap,He:2009ua,Kersten:2007vk,Ibarra:2010xw,Nemevsek:2012cd}. If one goes to HN mass scales of the order of the light neutrinos, new contributions to the seesaw neutrino masses should be taken into account (see for example \cite{Donini:2012tt}).   Probing the nature of neutrinos has been one of the most interesting and elusive tasks in modern physics. Experimentally, whether they are Dirac or Majorana fermions can be, in principle, established in neutrinoless double beta decay ($0\nu\beta\beta$) experiments \cite{Racah:1937qq,Furry:1939qr,Primakoff:1959chj,Primakoff:1970jy,Primakoff:1981sx,Schechter:1981bd,Doi:1985dx,Elliott:2004hr,Rodin:2007fz}, rare lepton number violating (LNV) decays of mesons \cite{Littenberg:1991ek,Littenberg:2000fg,Dib:2000wm,Ali:2001gsa,Ivanov:2004ch,deGouvea:2007qla,Delepine2011,LpezCastro2013,Abada:2013aba,Wang2014,Helo:2010cw,Atre:2009rg,Cvetic2010,Cveti2012,Cveti2014,Cveti2015,Milanes:2016rzr,Mandal2016,Zamora-Saa:2016qlk} and of $\tau$ lepton \cite{Gribanov2001,Cvetic:2002jy,Helo2011,ZamoraSaa2017,Zamora-Saa:2019naq}, and specific scattering processes \cite{Keung:1983uu,Tello2011,nemevsek2011neutrinoless,Kovalenko2009,Chen_2012,Chen_2013,Dev_2014,Das_2013,Das:2014jxa,Alva_2015,Das_2016,Das_2017,Degrande_2016,Das_2016b,Das:2017pvt,Buchmuller:1991tu,Kohda2013,Helo_2014,Dib_2015,Dib_2016,Dib_2017,Dib_2017a,Das_2017a,Das_2019a}.

The nature of Dirac neutrinos only allows them to appear in processes that are lepton number conserving (LNC). Majorana neutrinos can induce both lepton number conserving and lepton number violating (LNV) processes, which allows a wider spectrum of physics to take place.
An important example of this is baryogenesis via leptogenesis, where the LNV and CP-violating processes can lead to a generation of a lepton number asymmetry in the early universe, which is then converted (through sphaleron processes~\cite{tHooft:1976rip,tHooft:1976snw,Mohapatra:1979ia}) to the baryon number asymmetry observed in the universe~\cite{Ade:2015xua}. There are many different models that try to explain this asymmetry. However, two standard approaches that use Majorana neutrinos for successful Leptogenesis are \emph{out-of-equilibrium HN decays} (or \emph{Thermal Leptogenesis}) and  \emph{leptogenesis from oscillations}. Both of them use sterile neutrinos as an extension to the standard model, with their masses being calculated with the seesaw type-I mechanism. This mechanism allows us to have heavy neutrinos using the fact that the SM neutrinos have very low masses. These HNs satisfy the Sakharov conditions~\cite{Sakharov:1991} in order to produce the asymmetry dynamically. Consequently, thermal leptogenesis~\cite{Fukugita:1986hr,Buchmuller:2005eh,Buchmuller:2004nz} takes into account the lepton number asymmetry generated by the decay of a massive Majorana neutrino in a thermal bath, while the latter, known as Akhmedov-Rubakov-Smirnov (ARS) mechanism~\cite{Akhmedov:1998qx}, leads to a lepton number asymmetry by means of HN oscillations. The main difference between the two mechanisms comes from the fact that the first case is a \emph{freeze-out} situation while the ARS mechanism can be seen as a \emph{freeze-in} one.

The range of the HN masses for thermal leptogenesis is dictated by the amount of CP violation that can be generated\footnote{In the type-I seesaw mechanism, the mass scale was first discussed in \cite{Davidson:2002qv} and it is known as the \emph{Davidson-Ibarra bound}.}. In the most simple scenarios leptogenesis is constructed with masses $M_{N}\gtrsim 10^{10}$ GeV, or $M_{N}\gtrsim$ 1 TeV if one takes into account resonant effects~\cite{Pilaftsis:2003gt},  whereas the ARS mechanism allows neutrinos to reach masses as low as $\sim 1$ GeV. The HN mass scale for thermal leptogenesis cannot be reached in modern experiments, while ARS leptogenesis allows a variety of experiments to try and probe not only the nature of neutrinos, but also leptogenesis~\cite{Chun:2017spz}.

The search for the CP violation has been studied in different scenarios: resonant (overlap)  scattering processes~\cite{Pilaftsis:1997jf,Bray:2007ru,Hernandez:2018cgc}, resonant leptonic~\cite{Cvetic:2014nla,Dib:2014pga,Cvetic:2015naa} and semileptonic rare meson decays~\cite{Dib:2014pga,Cvetic:2013eza,Abada:2019bac}, as well as $B$ mesons, $W$ bosons and $\tau$ decays that include heavy neutrinos oscillation~\cite{Cvetic:2015ura,Cvetic:2018elt,Cvetic:2019rms,Zamora-Saa:2016ito,Zamora-Saa:2019naq,Anamiati:2016uxp,Antusch:2017ebe,Das:2017hmg}. The resonant (overlap) effect comes from the interference between two almost degenerate neutrino mass eigenstates with masses of order $\sim$ GeV. 

This article is organized in the following way: In Sec.~\ref{s2} we present the effective CP-violating $B$ meson decay width for the LNV process $B^{\pm} \to D^0 \ell^{\pm}_1 \ell^{\pm}_2 \pi^{\mp}$, and in Appendices \ref{sec:appA}-\ref{sec:appDW} more details are given. In Sec.~\ref{sec:simu} we present the numerical results for this effective branching ratio (with $\ell_1=\ell_2=\mu$) and for the related CP asymmetry ratio, for different values of the detector length, of the ratio of the HN mass difference and the HN total decay width, and for different values of the CP-violating phase. In Sec.~\ref{sec:dis} we discuss the possibility for the detection of various such signals within the detector at Belle II and summarize our results.

\section{CP Violation in Heavy Neutrino Decay}
\label{s2}

The simplest extension of the SM that explains the smallness of the active neutrino masses is the addition of right-handed neutrinos ($\nu_R$). Then, the relevant terms of the new Lagrangian $\mathcal{L}_N$ will read
\begin{equation}\label{lagranseesaw}
-\mathcal{L}_N=Y_{\nu}\overline{\ell}_L\phi\nu_R+ \frac{M_R}{2}\overline{(\nu_R)^c}\nu_R+h.c.\ ,
\end{equation}
where $M_R$ is the mass of the right-handed neutrinos. After diagonalizing the mass matrix, three very light neutrinos are obtained, as well as three heavy ones, this is the well known seesaw mechanism \cite{GellMann:1980vs,Sawada:1979dis,PhysRevLett.44.912}. The mass of the light neutrinos will be given by 
\begin{equation}
m_{\nu} \propto \langle\phi\rangle^2\frac{Y_{\nu}Y^T_{\nu}}{M_N}\ ,
\end{equation}
where $M_N=M_R$ is a $3 \times 3$ mass matrix of the heavy neutrinos and $\langle\phi\rangle$ is the electroweak vacuum expectation value of the Higgs field. By tuning the parameters in the above equation one can reach neutrino masses $\sim 1$ GeV, resulting in Yukawa couplings $\sim 10^{-5}$. This type of scenario is well discussed in the $\nu$MSM model~\cite{Asaka:2005pn,Asaka:2005an}. Two key ingredients in this model are the CP violation that occurs in the mixing of the heavy neutrinos and a resonant effect when the masses of two of them satisfy the condition $\Delta M_N$ ($\equiv M_{N_2}-M_{N_1}$) $ = \Gamma_N$.

\begin{figure}[H]
\centering
\includegraphics[scale = 0.85]{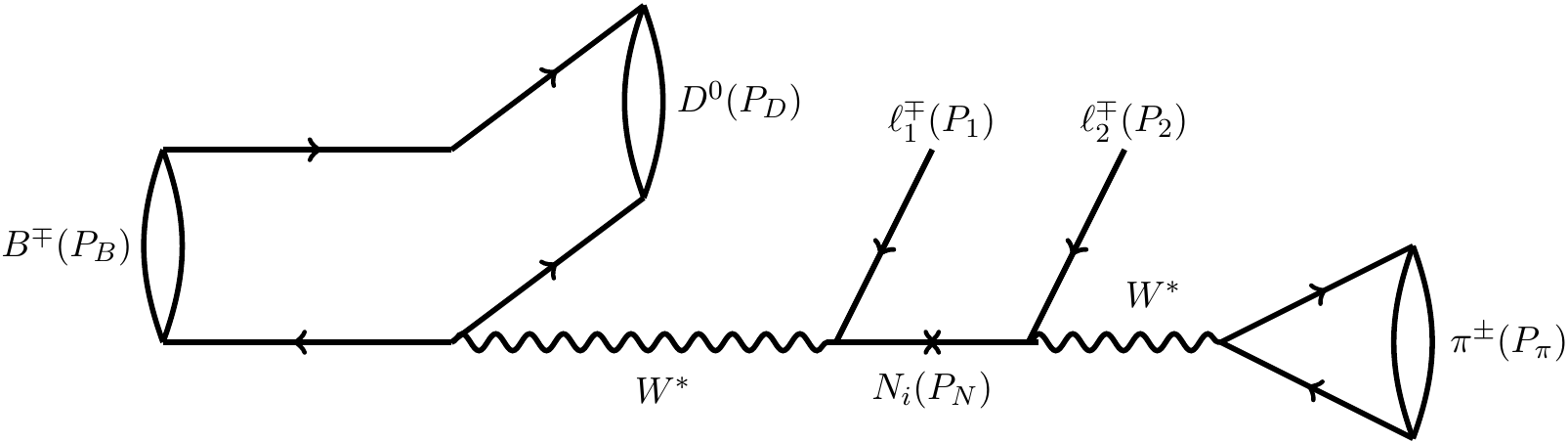}
\caption{Feynmann diagram of the decay processes.}
\label{fig:decproc}
\end{figure}

In previous articles we explored the HN CP-violating decays: $i)$ considering only resonant CP violation without HN oscillation effects \cite{Cvetic:2013eza,Cvetic:2014nla,Cvetic:2015naa,Zamora-Saa:2016ito} and $ii)$ nonresonant HN oscillation effects \cite{Cvetic:2015ura,Cvetic:2018elt,Cvetic:2019rms,Zamora-Saa:2019naq}. In this article, we will considerer the decay $B^{\pm} \to D^0 \ell^{\pm}_1 \ell^{\pm}_2 \pi^{\mp}$ (see Fig.~\ref{fig:decproc}) extending the previous analysis, by considering simultaneously both of the aforementioned CP-violating sources, in order to explore these signals at Belle II experiment. 

In this work we will assume the existence of several (three) Heavy Neutrino states $N_j$ ($j=1,2,3$), with respective masses $M_{N_j}$. In addition, we will assume that the first two heavy neutrinos are almost degenerate and with masses in the range of $\sim 1$ GeV, and the third neutrino is much heavier
\be
M_{N_3} \gg M_{N_2} \approx M_{N_1} \sim 1 \ {\rm GeV} \qquad (M_{N_2} > M_{N_1})\ .
\label{Masses}
\ee
The first three active neutrinos $\nu_{\ell}$ (where $\ell = e, \mu, \tau$) will have, in general, admixtures of the above mentioned heavy mass eigenstates
\be
\nu_{\ell} = \sum_{j=1}^{3} B_{\ell j} \nu_j + B_{\ell N_1} N_1 + B_{\ell N_2} N_2 + B_{\ell N_3} N_3\ ,
\ee
where the heavy-light mixing elements $B_{\ell N_j}$ are, in general, small complex numbers
\be
\label{mix}
B_{\ell_k N_j} \equiv |B_{\ell_k N_j}|e^{i \phi_{kj}}\ , \quad (k,j = 1,2,3) \ .
\ee

We will consider CP-violating decays of $B$ mesons into two light leptons ($\ell_1 \ell_2$) and a pion, mediated by heavy on-shell neutrinos $N_j$ ($j=1,2$). It turns out that (effective) branching ratios for the decays of the type $B \to D \ell_1 \ell_2 \pi$ (cf.~Fig.~\ref{fig:decproc}) are significantly larger than the decays $B \to \ell_1 \ell_2 \pi$, by about a factor of $30$-$40$ when $M_N \approx 2$ GeV, cf.~Ref.~\cite{Cvetic:2016fbv} (Figs.~19a and 20a there),\footnote{Majorana neutrinos in $B$ meson decays were considered also in Refs.~\cite{Cvetic:2017vwl,Duarte:2019rzs,Duarte:2020vgj}.} the main reason been the different CKM matrix elements $|V_{cb}| \sim 10 \; |V_{ub}|$. For this reason, we will consider the decay channels $B \to D \ell_1 \ell_2 \pi$, Fig.~\ref{fig:decproc}. The heavy neutrino $N_3$ will not enter our considerations because, in contrast to $N_1$ and $N_2$, it is off-shell in these decays. Furthermore, in order to avoid the kinematic suppression from heavy leptons, we exclude from our consideration the case of $\tau$-lepton production. In addition, to avoid the present stringent upper bounds on the heavy-light mixing $B_{e N_j}$, we also exclude from our consideration the case of $\ell=e$ lepton production. Thus, we will take $\ell_1 \ell_2 = \mu \mu$. The $N_1$-$N_2$ oscillation effects in such decays ($\ell_1 \ell_2 = \mu \mu$) turn out to disappear in LNC decays but survive in LNV decays \cite{Cvetic:2015ura}. Hence, we will consider the LNV decays $B^{\pm} \to D^0 \mu^{\pm} \mu^{\pm} \pi^{\mp}$, Fig.~\ref{fig:decproc}.  The CP-violating $B$ meson decay width for such a process, which accounts for the fact that the process will be detected only if the HN decays during its crossing through the detector ({\it effective\/} $\Gamma$), and includes both the overlap (resonant) \cite{Cvetic:2013eza,Cvetic:2014nla} and the HN-oscillation CP-violating sources \cite{Cvetic:2015ura,Cvetic:2018elt,Cvetic:2019rms}, is given by
\begin{align}
\label{DWeff}
\nonumber
\Gamma_{\rm eff}(B^{\pm} & \to D^0 \ell^{\pm}_1 \ell^{\pm}_2 \pi^{\mp}) =  \bG(B^{\pm} \to D^0 \ell^{\pm}_1 N) \ \bG(N \to \ell^{\pm}_2 \pi^{\mp}) 2 |B_{\ell_1 N}|^2  |B_{\ell_2 N}|^2  \\
\nonumber 
& \times \Bigg[\frac{1}{\Gamma_N}\Big(1- \exp \left( - \frac{-L \Gamma_N}{\gamma_N^{''} \beta_N^{''}} \right) \Big) \Big(1+\delta(Y) \cos(\theta_{LV})\mp \frac{\eta(Y)}{Y} \sin(\theta_{LV}) \Big) \\
\nonumber
& + \frac{1}{\Gamma_{N} (1+Y^2)}\Bigg\{
\exp \left( - \frac{-L \Gamma_N}{\gamma_N^{''} \beta_N^{''}} \right)
\Big[Y \sin\Big(\frac{2\pi L}{L_{\rm osc}}\pm \theta_{LV}\Big)-\cos\Big(\frac{2\pi L}{L_{\rm osc}}\pm \theta_{LV}\Big) \Big] \\
& + \Big(\cos(\theta_{LV}) \mp Y \sin(\theta_{LV}) \Big)\Bigg\} \Bigg]\ ,
\end{align}
where $L$ stands for the distance (in the lab frame) between the two vertices of the process (the flight length of the on-shell neutrino $N_j$),\footnote{$L$ is thus limited by the (effective) length $L_{\rm det}$ of the detector, $L \leq L_{\rm det}$. The lab frame in this work is denoted by $\Sigma^{''}$. However, for simplicity of notation, the distance $L^{''}$ in the lab frame will be denoted simply as $L$.}
  $L_{\rm osc}=(2\pi\gamma_N^{''} \beta_N^{''})/\Delta M_N$ is the HN oscillation length,
\be
Y \equiv \frac{ \Delta M_N }{\Gamma_N}\ , \qquad \Delta M_N \equiv M_{N_2}-M_{N_1} (> 0)\ ,
\label{Ydef} \ee 
and $\theta_{LV}$ is the CP-violating phase\footnote{For example, if $\ell_1 = \ell_2 = \mu$, then $\theta_{LV} \equiv \theta_{21} = 2(\phi_{\mu 2}-\phi_{\mu 1})=2 [{\rm arg}(B_{\mu N_2})-{\rm arg}(B_{\mu N_1})]$.} which, according to the notation of Eq.~\eqref{mix} can be written as
\be
\theta_{LV} \equiv \theta_{kj} = (\phi_{1k}+\phi_{2k}-\phi_{1j}-\phi_{2j})\ , \quad (k,j = 1,2)\ .
\ee
Further, the functions $\eta(Y)$ and $\delta(Y)$ are \cite{Cvetic:2013eza,Cvetic:2014nla,Cvetic:2015naa}
\be
\eta(Y) = \frac{Y^2}{Y^2+1}\ , \qquad \delta(Y)=\frac{1}{Y^2+1}\ .
\label{etadel} \ee 
The numerical values of $\eta(Y)$ and $\delta(Y)$ were obtained in \cite{Cvetic:2013eza,Cvetic:2014nla}, and the explicit expression for $\eta(Y)$ was obtained in \cite{Cvetic:2015naa} (App.~6 there). Based on the mentioned numerical values of $\delta(Y)$  (cf.~Table I in \cite{Cvetic:2013eza}, Table II in \cite{Cvetic:2014nla}, and Table 4 in \cite{Cvetic:2015naa}), we observe a posteriori here that they can be reproduced with high precision by the explicit expression for $\delta(Y)$ given here. The functions $\delta(Y)$ and $\eta(Y)$ are related with the real and imaginary parts, respectively, of the product of scattering amplitudes for the processes $W^{*} \to \ell_1 N_j \to \ldots $ ($j=1,2$), and they involve the product of (almost on-shell) propagators of the nearly degenerate neutrinos $N_1$ and $N_2$. We refer for details to Refs.~\cite{Cvetic:2013eza,Cvetic:2014nla,Cvetic:2015naa}.

In Eq. \eqref{DWeff}, the HN Lorentz kinematical parameters in the lab frame ($\Sigma^{''}$)  $\beta_N^{''}$ and $\gamma_N^{''} = 1/\sqrt{1 - (\beta_N^{''})^2}$ are assumed to be constant. This can be extended to the realistic case of variable $\beta_N^{''}$ \cite{Cvetic:2017vwl}, and this extension is explained in Appendix \ref{sec:appC}. We also assumed that $|B_{\ell N_1}|=|B_{\ell N_2}|$ ($\equiv |B_{\ell N}|$), with $\ell=\mu, e, \tau$.

Furthermore, the expression (\ref{DWeff}), in addition to the aforementioned approximations (fixed $\beta_N^{''}$ and common $|B_{\ell N}|$'s), is obtained in an approximation of combining the overlap (resonant) and oscillation effects, which is valid when $Y$ is significantly larger than one, e.g.~ $Y \gtrsim 5$. This is explained in more detail in Appendix \ref{sec:appDW}, where several steps of derivation of the expression (\ref{DWeff}) are given.

\begin{figure}[H]
\centering
\includegraphics[scale = 0.65]{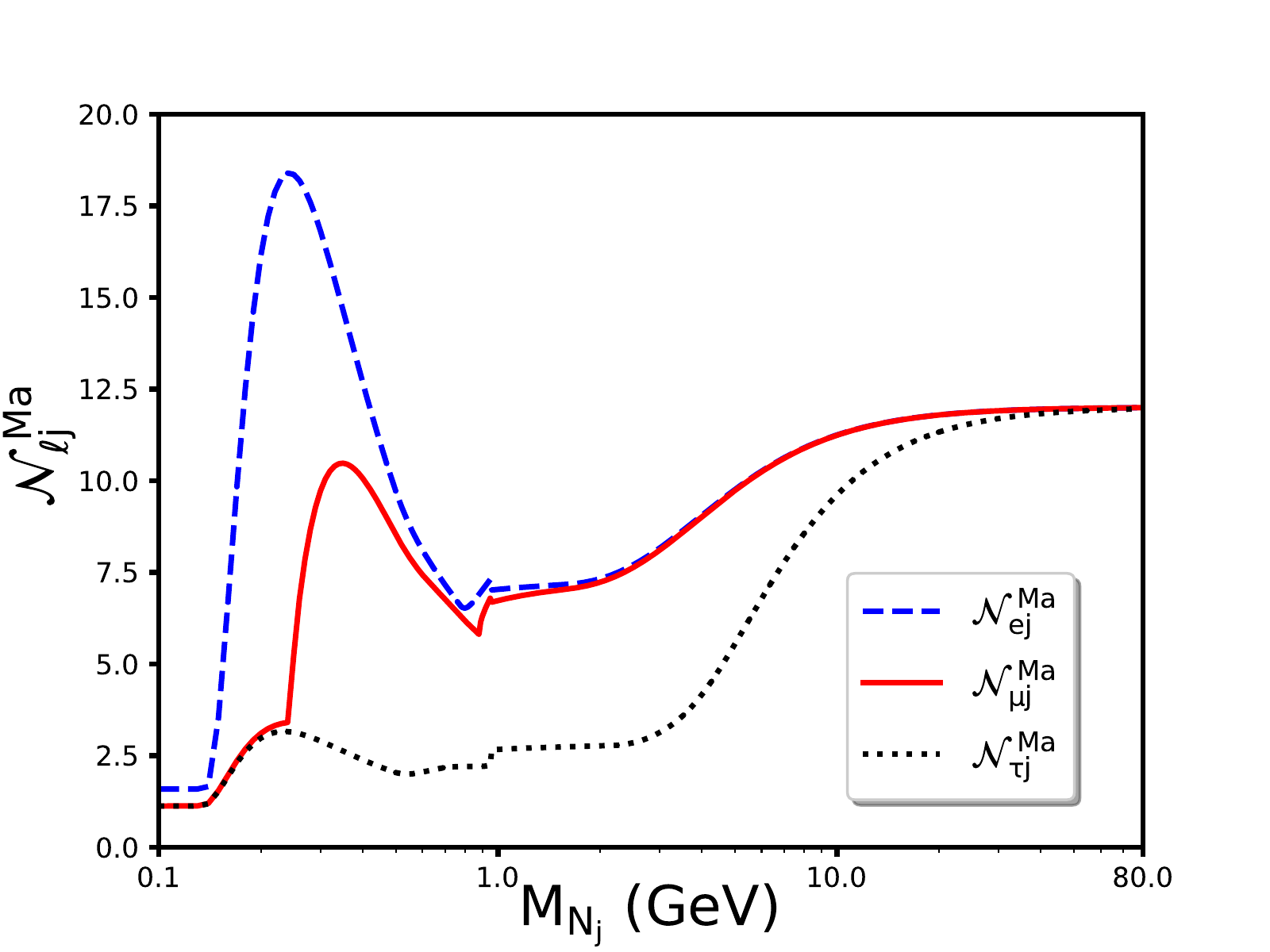}
\caption{Effective mixing coefficients ${\cal N}_{\ell j}^{\rm Ma}$ for Majorana neutrinos.}
\label{fig:efcoef}
\end{figure}

In general, $\Gamma_N = (\Gamma_{N_1} + \Gamma_{N_2})/2$ where $\Gamma_{N_j}$ is the total decay width of HN $N_j$ ($j=1,2$). However, due to our assumption $|B_{\ell N_1}|=|B_{\ell N_2}|$ ($\equiv |B_{\ell N}|$), we have $\Gamma_{N_1} = \Gamma_{N_2} = \Gamma_N$. This is because the total decay width of the heavy neutrino $N_j$ is \cite{Cvetic:2014nla,Cvetic:2015naa}
\begin{equation}
\Gamma_{N_j}   \approx  \K_j^{\rm Ma}\ \frac{G_F^2 M_{N_j}^5}{96\pi^3} \ , \qquad
  \K_i^{\rm Ma} = {\cal N}_{e j}^{\rm Ma} \; |B_{e N_j}|^2 + {\cal N}_{\mu j}^{\rm Ma} \; |B_{\mu N_j}|^2 + {\cal N}_{\tau j}^{\rm Ma} \; |B_{\tau N_j}|^2\ ,
\label{DNwidth}
\end{equation}
where ${\cal N}_{\ell j}^{\rm Ma}$ are the effective mixing coefficients whose range is $\sim 1$-$10$ and account for all possible HN decay channels. The $N_j$ coefficients are presented in Fig.~\ref{fig:efcoef}.

From now on, as mentioned earlier we will consider only the case $\ell_1 = \ell_2 = \mu$. We notice that $|B_{\mu N_j}|^2 \approx |B_{\tau N_j}|^2 \lesssim 10^{-5}$ and $|B_{e N_j}|^2<10^{-7}$, so that the $\K_j^{\text{Ma}}$ can receive significant contribution only from $\mu$ and $\tau$ decay channels (note that ${\cal N}_{\mu j}^{\rm Ma} + {\cal N}_{\tau j}^{\rm Ma} \approx 10$). The mixings $B_{\mu N_1}$ and $B_{\mu N_2}$ can be, in principle, significantly different for the two HNs, and therefore, the two mixing factors $\mathcal{K}^{\rm Ma}_j$ $(j = 1, 2)$ may differ significantly from each other. However, as mentioned earlier, in this work we will assume that $|B_{\ell N_1}|=|B_{\ell N_2}|$ ($\equiv |B_{\ell N}|$). Taking $\mathcal{K}^{\rm Ma}_1 \approx \mathcal{K}^{\rm Ma}_2= 10 \; |B_{\mu N}|^2$ the HN total decay width then reads
\begin{equation}
\label{DNwidthappr}
\Gamma_N(M_N)  = 10 \; |B_{\mu N}|^2 \ \frac{G_F^2 M_{N}^5}{96\pi^3}\ ,
\end{equation}
we also note that the HN masses are almost equal, i.e. $M_{N_j} \simeq M_N$.

The usual measure of the relative CP violation effect is given by the CP asymmetry ratio 
\be
A_{\rm CP} = \frac{\Gamma_{\rm eff}(B^{+}  \to D^0 \mu^{+} \mu^{+} \pi^{-})-\Gamma_{\rm eff}(B^{-}  \to D^0 \mu^{-} \mu^{-} \pi^{+})}{\Gamma_{\rm eff}(B^{+}  \to D^0 \mu^{+} \mu^{+} \pi^{-})+\Gamma_{\rm eff}(B^{-}  \to D^0 \mu^{-} \mu^{-} \pi^{+})} \ .
\label{acp}
\ee
\section{results}
\label{sec:simu}
In this Section we show the numerical results for the effective branching ratio ${\rm Br_{eff}(B^{\pm})}=\Gamma_{\rm eff}(B^{\pm}  \to D^0 \mu^{\pm} \mu^{\pm} \pi^{\mp})/\Gamma(B\to \text{all})$ and the CP asymmetry ratio ${\rm A_{CP}}$ in \eqref{acp} for different values of the $Y$ parameter and the maximal displaced vertex length $L$, which can be interpreted as the (effective) detector length ($L \leq L_{\rm det}$). The calculations were performed by numerical integration with the VEGAS algorithm \cite{Lepage:1977sw} in each step of $L$ and $Y$. All integrations were performed using $M_N=2$ GeV and heavy-light mixings $|B_{\mu N}|^2=|B_{\tau N}|^2=10^{-5}$. The selected mixing values are consistent with the present experimental constraints given in Ref.~\cite{Atre:2009rg,Abada:2017jjx} and references therein. Moreover, two different values (scenarios) were chosen for the CP-violating phase: $\theta_{LV}=\pi/2,\pi/4$.

The kinematical Lorentz factor $\gamma_N^{''} $ and $\beta_N^{''}$ in Eq. \eqref{DWeff} in reality are not fixed, but vary and are obtained as explained in Appendix \ref{sec:appC} [Eq.~(\ref{bNgNpp})], where the general expression $\Gamma_{\rm eff}$ for the case of only one HN $N$ is given in Eq.~(\ref{Geff}). In the case of two (almost degenerate) HNs $N_j$ ($j=1,2$) the expression (\ref{Geff}) gets extended by the overlap (resonant) and oscillation terms as those appearing in Eq.(\ref{DWeff}), leading to our main formula
\bea
\label{DWeffgen}
\lefteqn{
\Gamma_{\rm eff}(B^{\pm}  \to D^0 \ell^{\pm}_1 \ell^{\pm}_2 \pi^{\mp}) =  2 |B_{\ell_1 N}|^2  |B_{\ell_2 N}|^2  \frac{ \bG(N \to \ell_2 \pi) }{\Gamma_N} 
\int d q^2 \int d \Omega_{{\hat q}'} \int d \Omega_{{\hat p}_1}
\frac{d \bG(B \to D \ell_1 N)}{ d q^2 d \Omega_{{\hat q}'}  d \Omega_{{\hat p}_1}}
}
\nonumber\\
&& \times \Bigg[ \left\{ 1 - \exp \left(- L \Gamma_N/\sqrt{ \left(E''_N(q^2;{\hat q}',{\hat p}_{1})/M_N \right)^2 - 1 } \right) \right\}
\Big[ 1+\delta(Y) \cos(\theta_{LV})\mp \frac{\eta(Y)}{Y} \sin(\theta_{LV}) \Big] \nonumber\\
&& + \frac{1}{(1+Y^2)} \Bigg\{
\exp \left(- L \Gamma_N/\sqrt{ \left(E''_N(q^2;{\hat q}',{\hat p}_{1})/M_N \right)^2 - 1 } \right)
\Big[Y \sin\Big(\frac{2\pi L}{L_{\rm osc}}\pm \theta_{LV}\Big)-\cos\Big(\frac{2\pi L}{L_{\rm osc}}\pm \theta_{LV}\Big) \Big]
\nonumber\\
&& + \Big(\cos(\theta_{LV}) \mp Y \sin(\theta_{LV}) \Big)\Bigg\} \Bigg] \ .
\eea 
Here we should keep in mind that the oscillation length $L_{\rm osc}$ relies on the (variable) Lorentz factors $\beta_N^{''}$ and $\gamma_N^{''}$, namely $2 \pi/L_{\rm osc}=Y \Gamma_N/(\gamma_N^{''} \beta_N^{''})$ [cf.Eq.~(\ref{Losc})], so it also depends on the integration variables $q^2$, ${\hat q}'$ and ${\hat p}_{1}$ via $E''_N(q^2;{\hat q}',{\hat p}_{1})$, cf.~Eq.~(\ref{bNgNpp}).\footnote{From the expression (\ref{DWeffgen}), and using Eqs.~(\ref{etadel}), it can be checked after some algebra that in the limit $Y=0$ the two decay widths (i.e., for $B^+$ and $B^-$) become equal to each other.}

In order to evaluate the relevance of Oscillatory and Overlapping effects on the main decay channel, we can either: (a) disregard in Eq.~\eqref{DWeffgen} the overlap (resonant) terms and include only the oscillatory terms
\bea
\label{osccontri}
\lefteqn{
\Gamma^{\rm osc}_{\rm eff}(B^{\pm}  \to D^0 \ell^{\pm}_1 \ell^{\pm}_2 \pi^{\mp}) =  2 |B_{\ell_1 N}|^2  |B_{\ell_2 N}|^2  \frac{ \bG(N \to \ell_2 \pi) }{\Gamma_N} 
\int d q^2 \int d \Omega_{{\hat q}'} \int d \Omega_{{\hat p}_1}
\frac{d \bG(B \to D \ell_1 N)}{ d q^2 d \Omega_{{\hat q}'}  d \Omega_{{\hat p}_1}}
}
\nonumber\\
&& \times \Bigg[ \left\{ 1 - \exp \left(- L \Gamma_N/\sqrt{ \left(E''_N(q^2;{\hat q}',{\hat p}_{1})/M_N \right)^2 - 1 } \right) \right\}
 \nonumber\\
&& + \frac{1}{(1+Y^2)} \Bigg\{
\exp \left(- L \Gamma_N/\sqrt{ \left(E''_N(q^2;{\hat q}',{\hat p}_{1})/M_N \right)^2 - 1 } \right)
\Big[Y \sin\Big(\frac{2\pi L}{L_{\rm osc}}\pm \theta_{LV}\Big)-\cos\Big(\frac{2\pi L}{L_{\rm osc}}\pm \theta_{LV}\Big) \Big]
\nonumber\\
&& + \Big(\cos(\theta_{LV}) \mp Y \sin(\theta_{LV}) \Big)\Bigg\} \Bigg] \ ;
\eea
(b) or we can disregard in Eq.~\eqref{DWeffgen} the oscillatory terms and include only the overlap (resonant) terms
\bea
\label{overcontri}
\lefteqn{
\Gamma^{\rm overlap}_{\rm eff}(B^{\pm}  \to D^0 \ell^{\pm}_1 \ell^{\pm}_2 \pi^{\mp}) =  2 |B_{\ell_1 N}|^2  |B_{\ell_2 N}|^2  \frac{ \bG(N \to \ell_2 \pi) }{\Gamma_N} 
\int d q^2 \int d \Omega_{{\hat q}'} \int d \Omega_{{\hat p}_1}
\frac{d \bG(B \to D \ell_1 N)}{ d q^2 d \Omega_{{\hat q}'}  d \Omega_{{\hat p}_1}}
}
\nonumber\\
&& \times \Bigg[ \left\{ 1 - \exp \left(- L \Gamma_N/\sqrt{ \left(E''_N(q^2;{\hat q}',{\hat p}_{1})/M_N \right)^2 - 1 } \right) \right\}
\Big[ 1+\delta(Y) \cos(\theta_{LV})\mp \frac{\eta(Y)}{Y} \sin(\theta_{LV}) \Big] \Bigg] \ .
\eea 

\begin{figure}[H]
\centering
\includegraphics[width=0.49\textwidth]{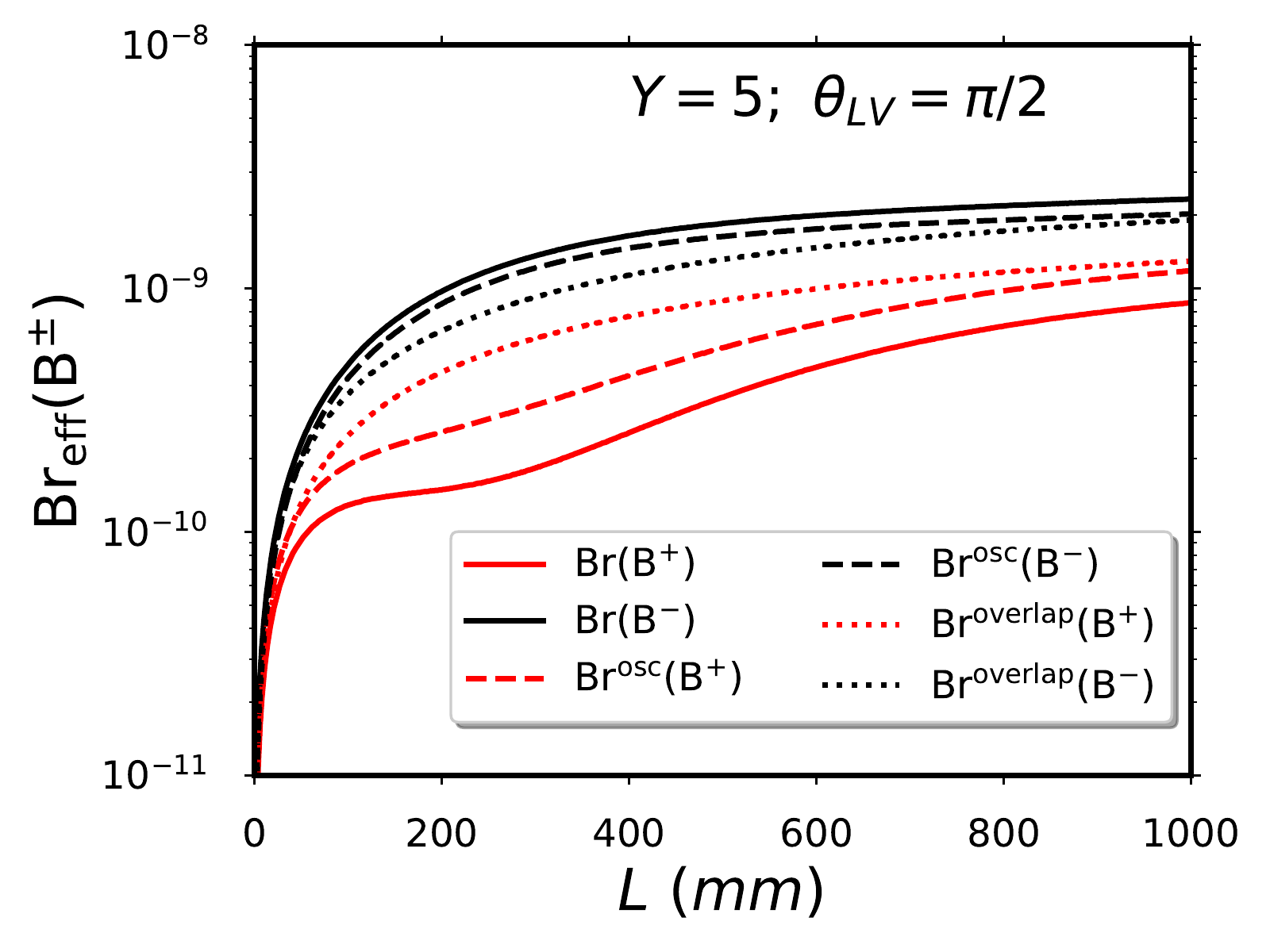}
\includegraphics[width=0.49\textwidth]{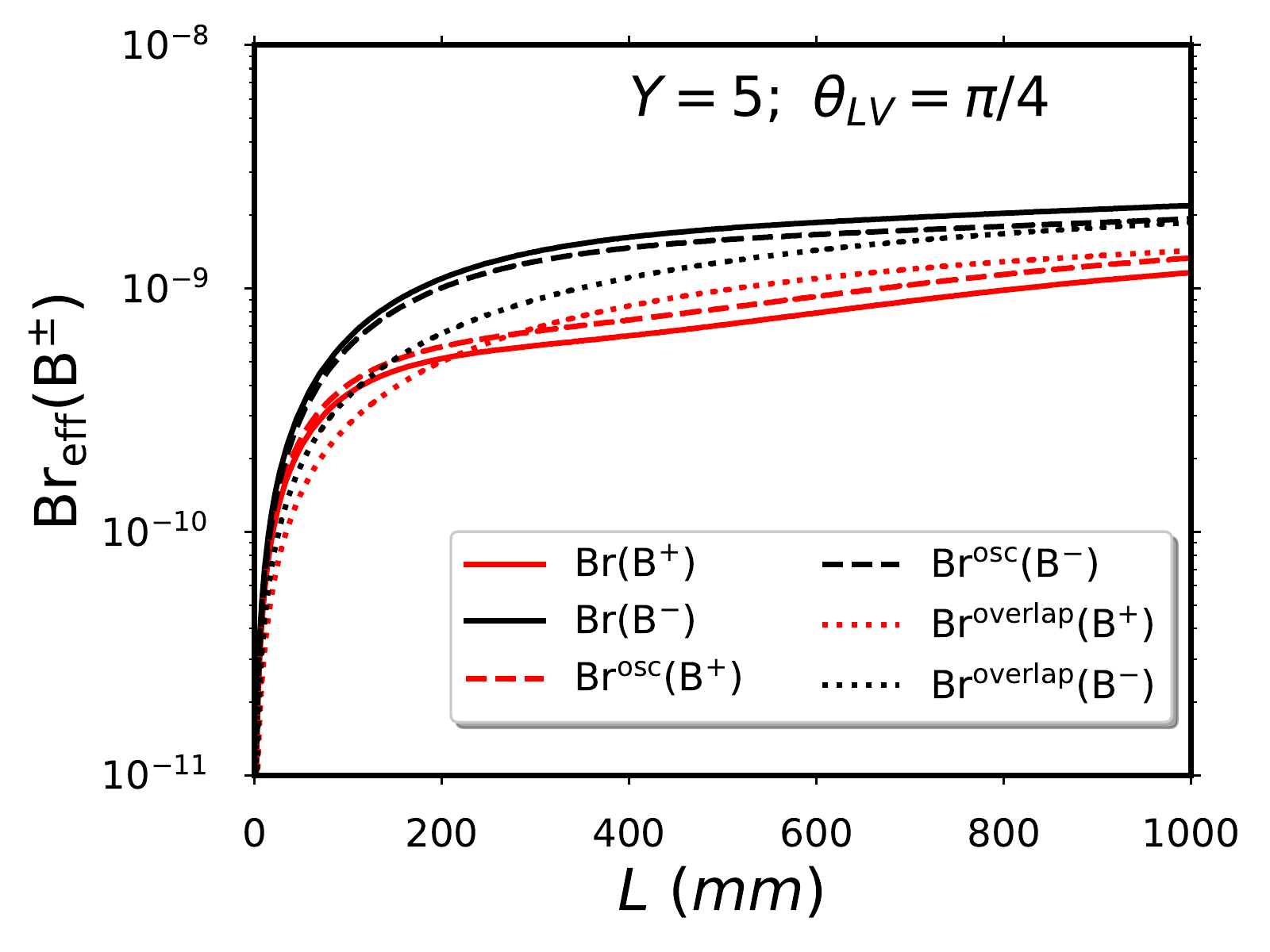}
\includegraphics[width=0.49\textwidth]{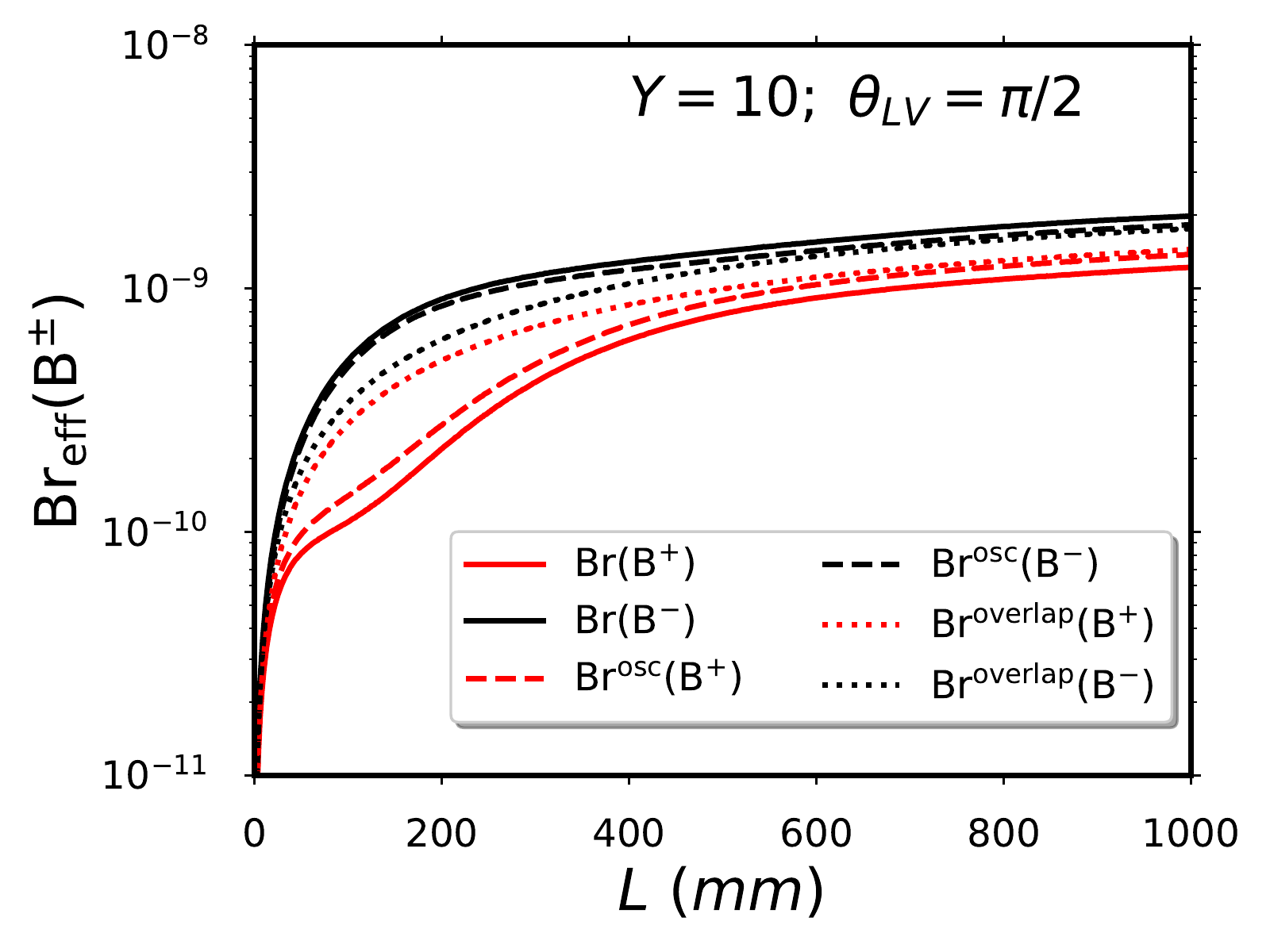}
\includegraphics[width=0.49\textwidth]{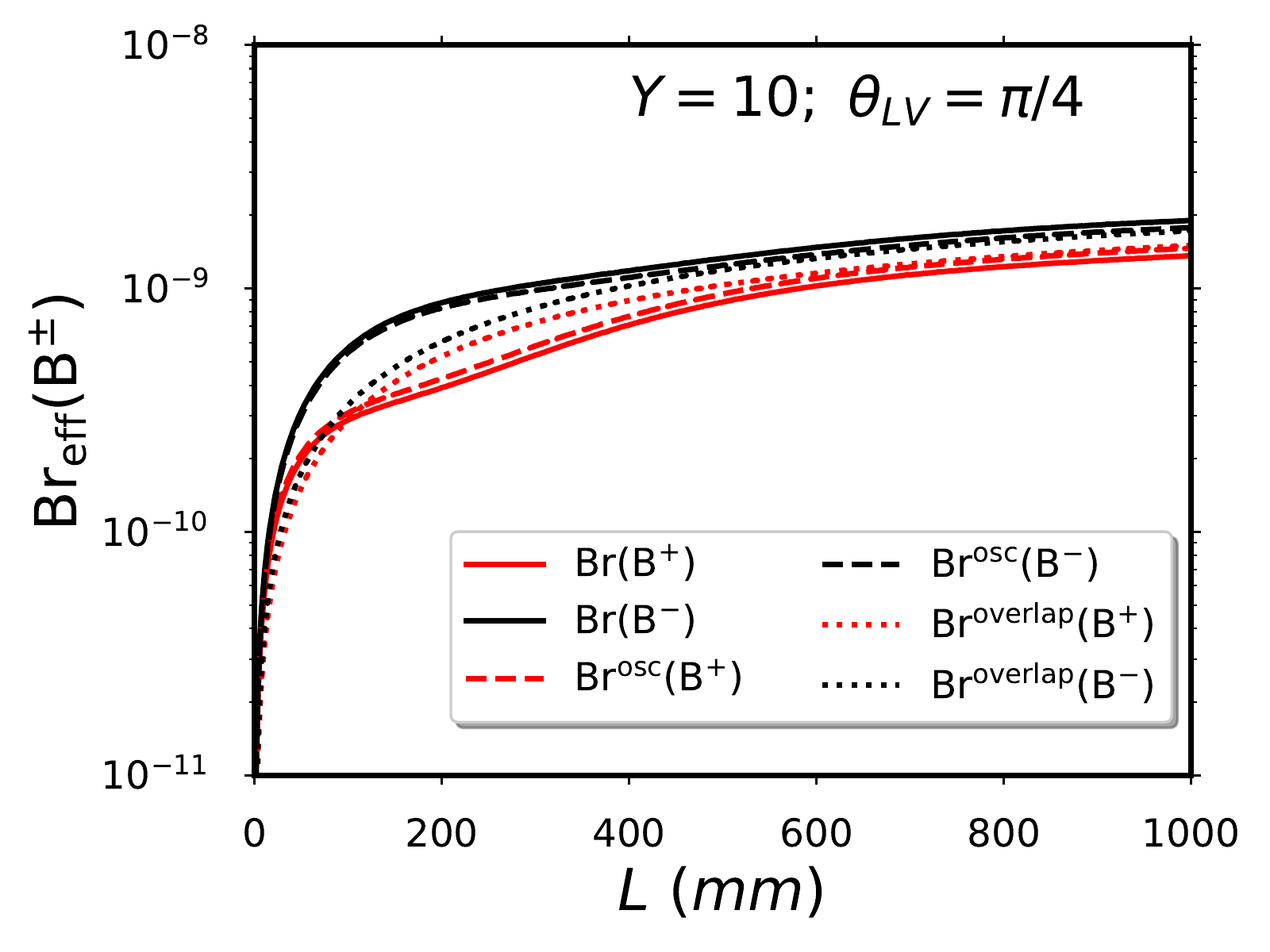}
\caption{Different contributions to the total effective branching ratio as a function of the maximal displaced vertex length ($L$) for $M_N=2.0$ GeV , $|B_{\mu N}|^2=|B_{\tau N}|^2=10^{-5}$ and different values of $Y$ and $\theta_{LV}$. Full lines stand for the case of $\Gamma_{\rm eff}(B^{\pm})$ (Eq.~\ref{DWeffgen}), dashed lines stand for $\Gamma^{\rm osc}_{\rm eff}(B^{\pm})$  (Eq.~\ref{osccontri}) and dotted ones for $\Gamma^{\rm overlap}_{\rm eff}(B^{\pm})$ (Eq.~\ref{overcontri}). The curve for $B^-$ are higher than those for $B^+$. }
\label{fig:compar}
\end{figure}

Figures \ref{fig:compar} show a comparison between Eqs.~\ref{DWeffgen}, \ref{osccontri} and \ref{overcontri}, as a function of the maximal displaced vertex length (effective detector length) $L$. We recall that the effective branching ratio is ${\rm Br}_{\rm eff} = \Gamma_{\rm eff}/\Gamma_B$, where $\Gamma_B=4.017 \times 10^{-13}$ GeV.
We can deduce from these figures that the oscillation contributions are usually larger in magnitude than the overlap (resonant) contributions, and that this trend gets stronger when $Y$ increases.

On the other hand, we notice that Figures \ref{fig:compar} show very small values of $\rm Br_{eff}$ when the detector length $L\approx 0$, this is consequent with the fact that at short distances only few neutrinos have decayed. On the contrary, for large $L$ all neutrinos have decayed, therefore the $\rm Br_{eff}$ becomes constant. In the expression Eq.~\eqref{DWeffgen} this situation is reflected when $L$ is so large that ${\rm exp}(-L \Gamma_N/(\gamma_N^{``} \beta_N^{''}))$ is almost zero, and consequently the oscillation contributions disappear and the $L$-dependence disappears.


We remark that both effects, oscillatory (Eq.~\ref{osccontri}) and overlap (Eq.~\ref{overcontri}), depend explicitly on $Y$. Therefore, it is relevant to explore how the Effective Branching Ratio changes while $Y\equiv \Delta_{M_N}/\Gamma_{N}$ varies for a fixed value of $L$.

\begin{figure}[H]
\centering
\includegraphics[width=0.49\textwidth]{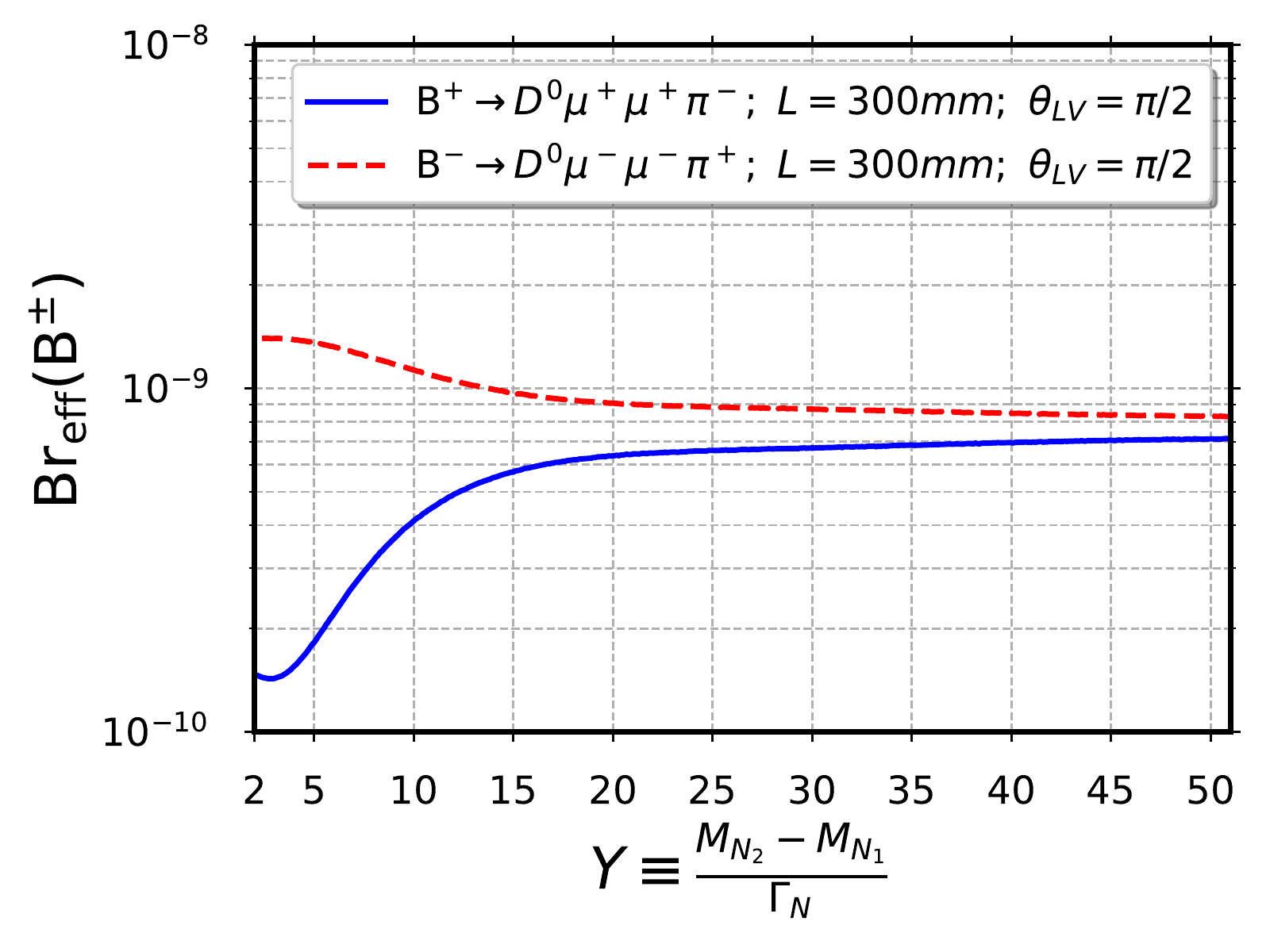}
\includegraphics[width=0.49\textwidth]{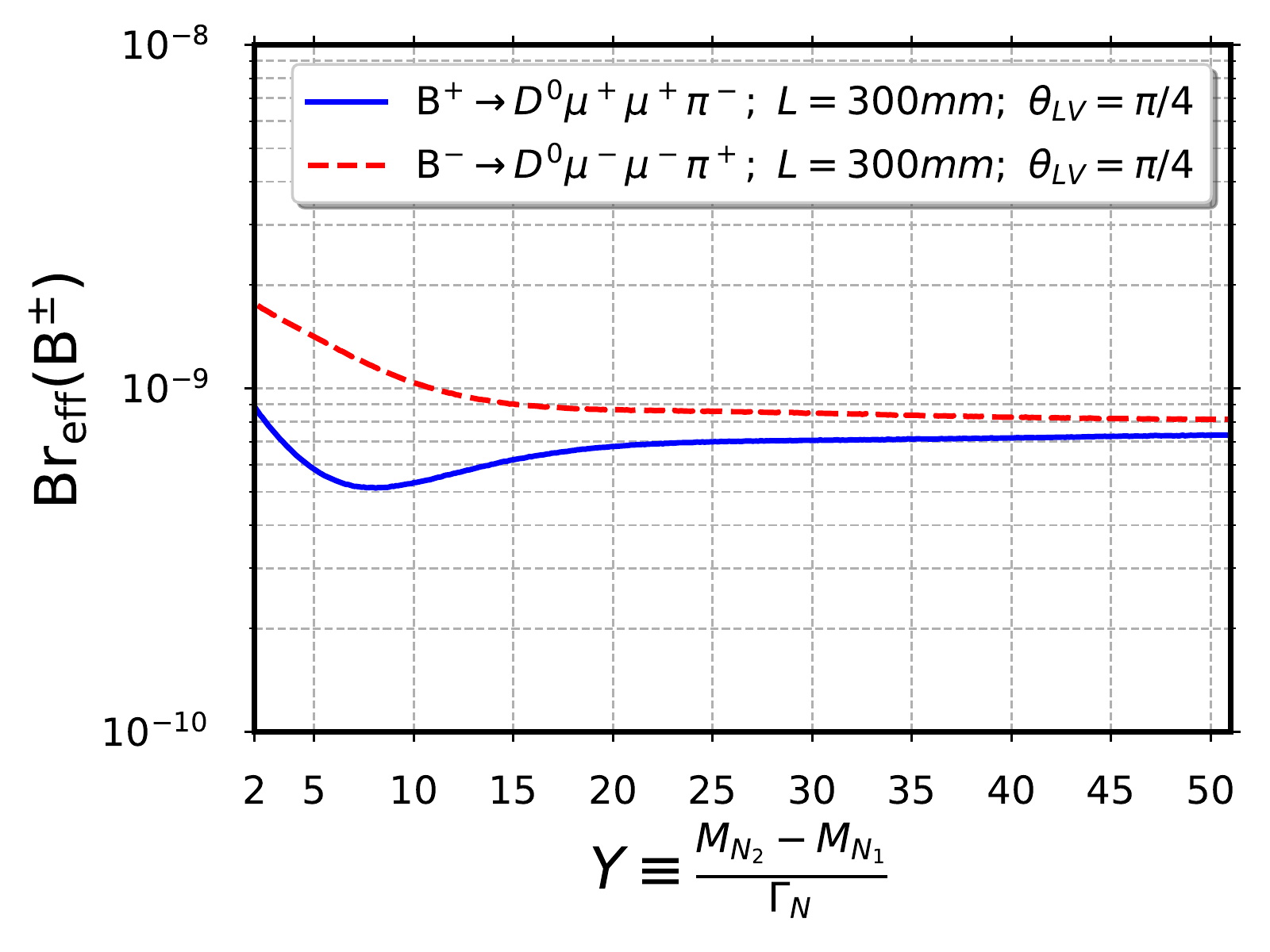}
\caption{Effective branching ratio as a function of $Y$ for $M_N=2.0$ GeV, $L=300$ mm and $|B_{\mu N}|^2=|B_{\tau N}|^2=10^{-5}$. Left Panel: $\theta_{LV}=\pi/2$. Right panel: $\theta_{LV}=\pi/4$.}
\label{fig:beff3}
\end{figure}

\begin{figure}[H]
\centering
\includegraphics[width=0.49\textwidth]{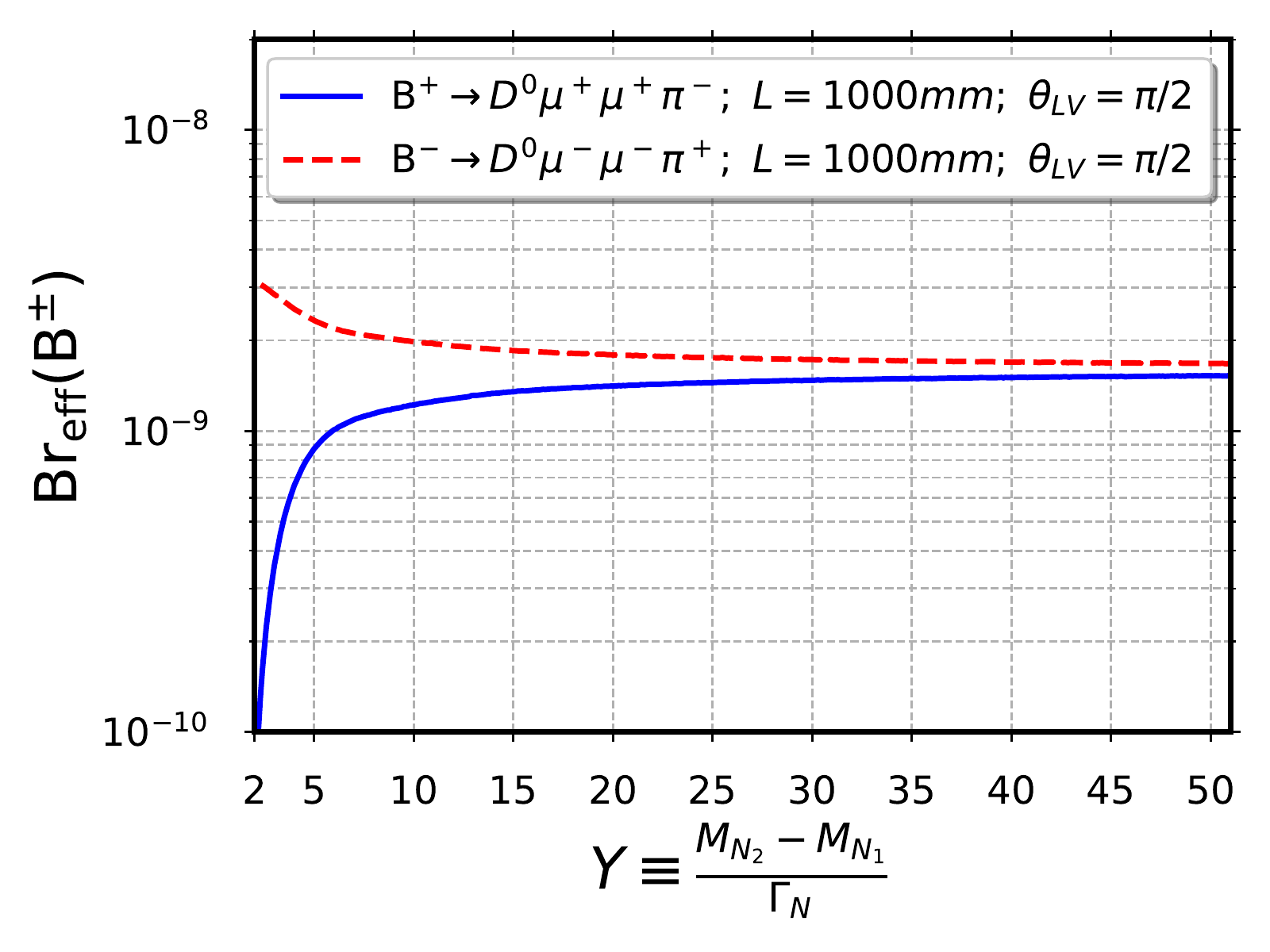}
\includegraphics[width=0.49\textwidth]{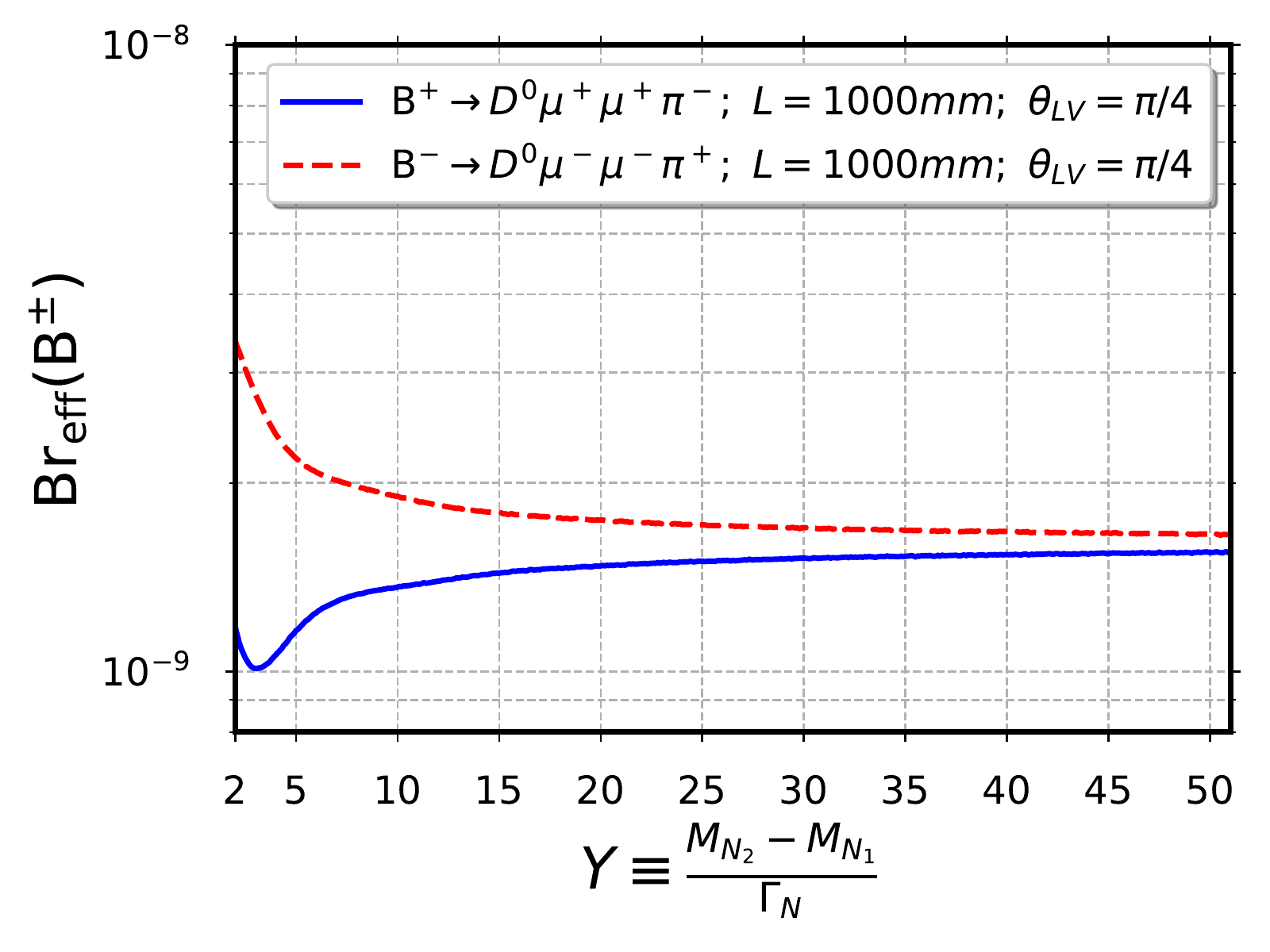}
\caption{Effective branching ratio as a function of $Y$ for $M_N=2.0$ GeV, $L=1000$ mm and $|B_{\mu N}|^2=|B_{\tau N}|^2=10^{-5}$. Left Panel: $\theta_{LV}=\pi/2$. Right panel: $\theta_{LV}=\pi/4$. }
\label{fig:beff4}
\end{figure}

Figures \ref{fig:beff3} and \ref{fig:beff4} show the effective branching ratio as a function of $Y$ for different fixed maximal displaced vertex lengths (effective detector lengths) $L=300$ mm and $L=1000$ mm, respectively.

\begin{figure}[H]
\centering
\includegraphics[width=0.49\textwidth]{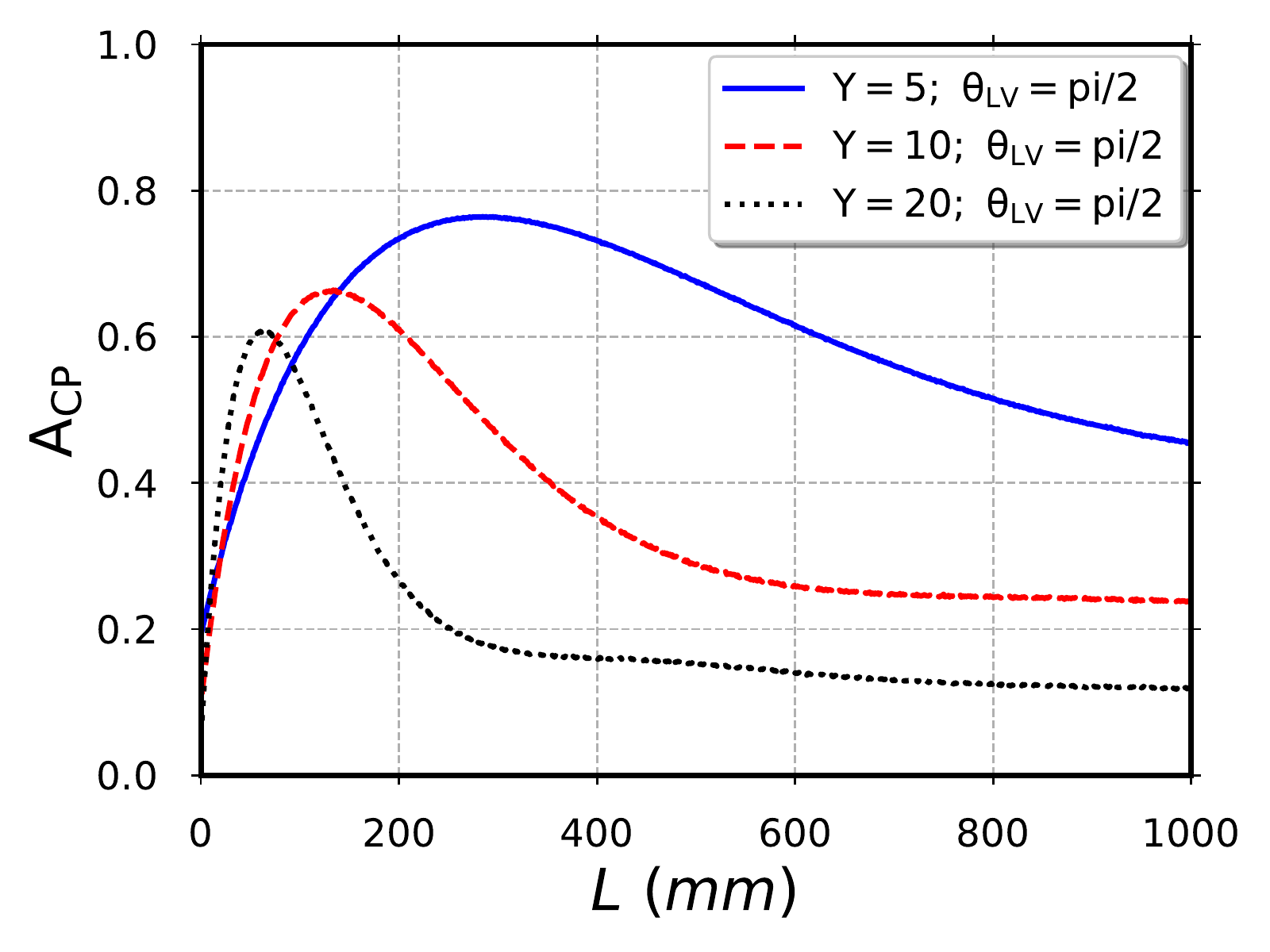}
\includegraphics[width=0.49\textwidth]{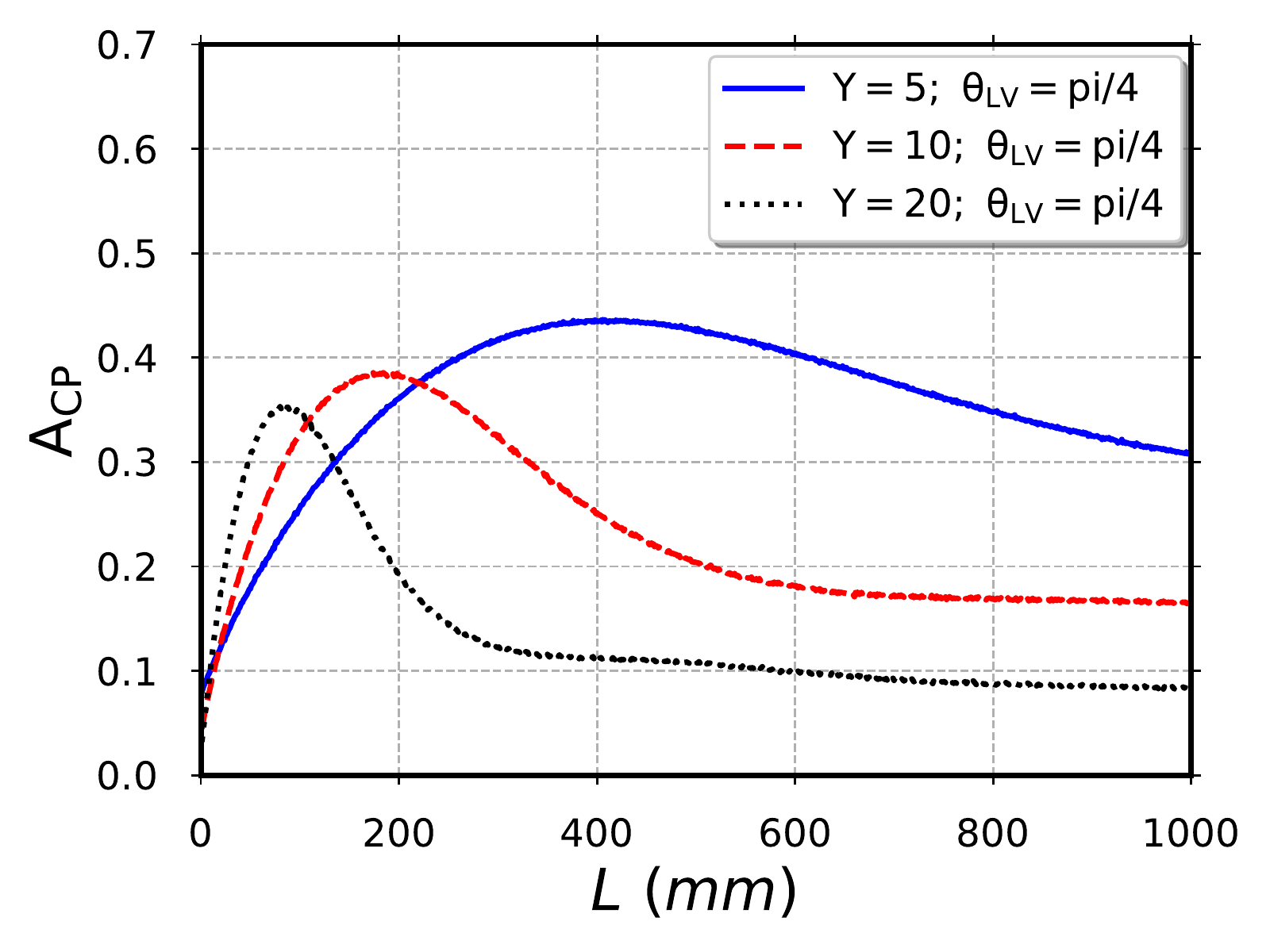}
\caption{CP asymmetry as a function of the maximal displaced vertex length ($L$) for $M_N=2.0$ GeV and $|B_{\mu N}|^2=|B_{\tau N}|^2=10^{-5}$. Left Panel: $\theta_{LV}=\pi/2$. Right panel: $\theta_{LV}=\pi/4$. }
\label{fig:acp1}
\end{figure}

\begin{figure}[H]
\centering
\includegraphics[width=0.49\textwidth]{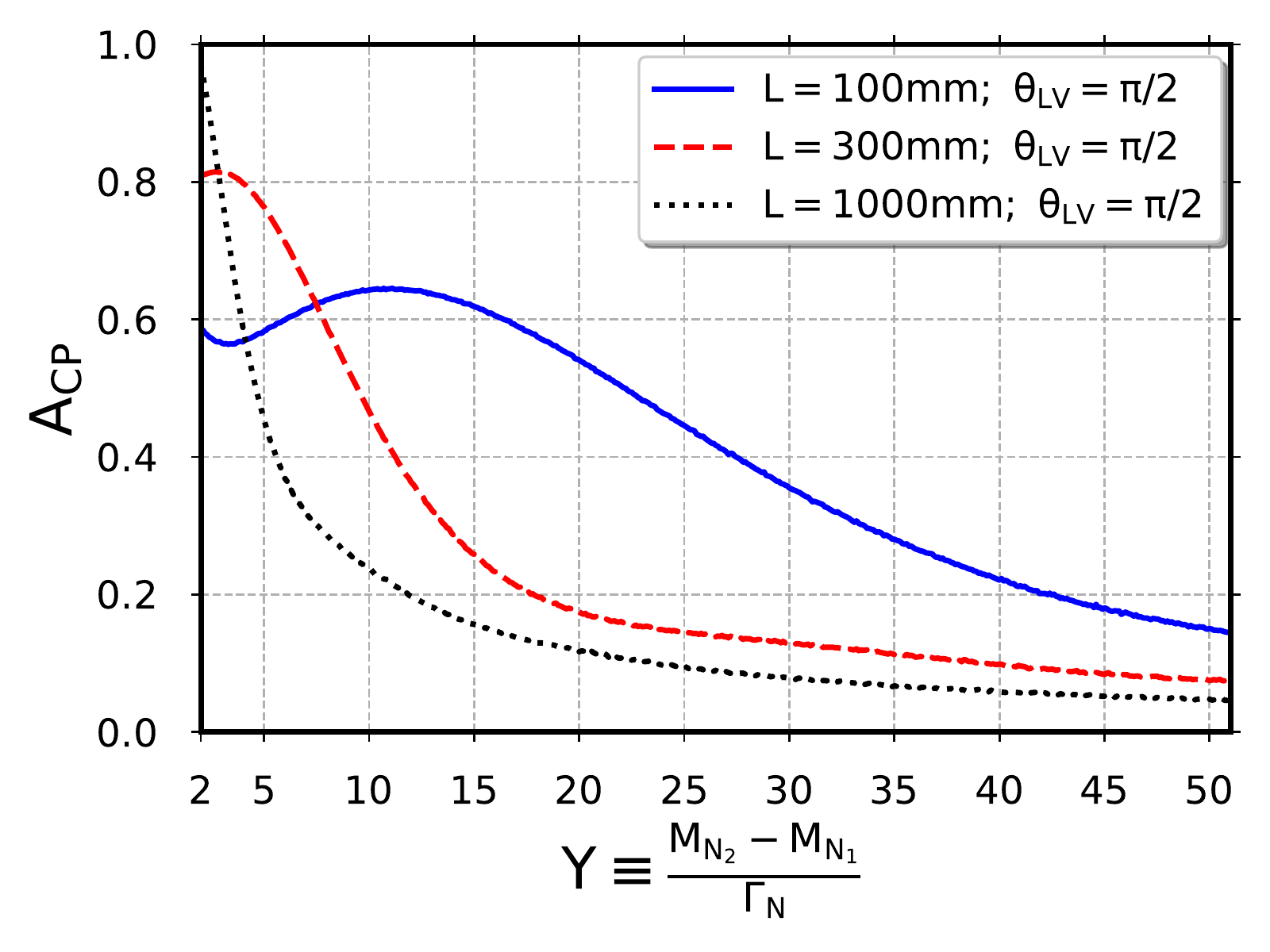}
\includegraphics[width=0.49\textwidth]{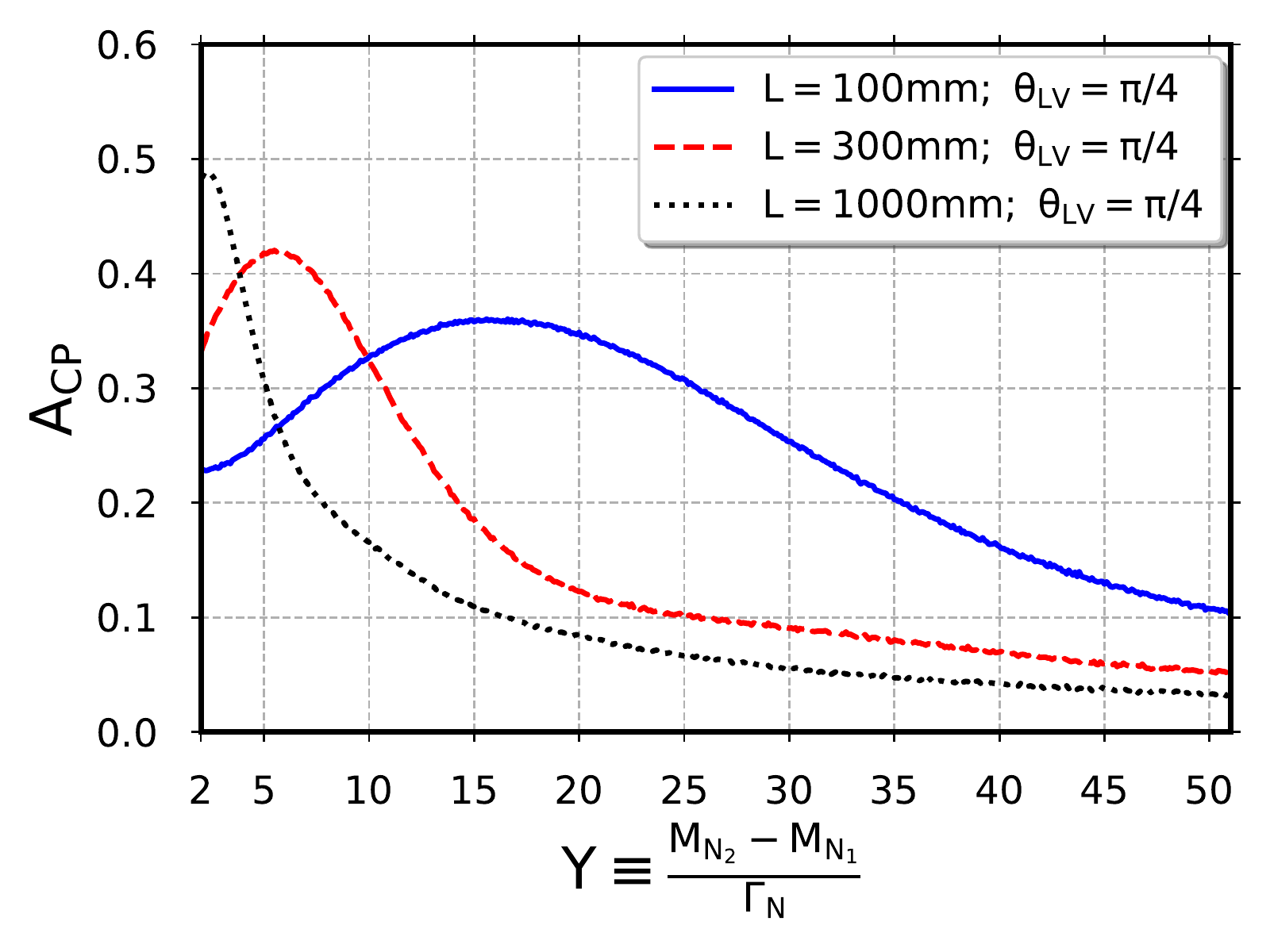}
\caption{CP asymmetry as a function of $Y$ for $M_N=2.0$ GeV and $|B_{\mu N}|^2=|B_{\tau N}|^2=10^{-5}$. Left Panel: $\theta_{LV}=\pi/2$. Right panel: $\theta_{LV}=\pi/4$. }
\label{fig:acp2}
\end{figure}

 Figure \ref{fig:acp1} shows the CP asymmetry ($\rm A_{CP}$) as a function of the maximal displaced vertex length $L$ for three different values $Y$. Figure \ref{fig:acp2} shows the CP asymmetry as a function of $Y$ for three different values of length $L$.

\begin{figure}[H]
\centering
\includegraphics[width=0.49\textwidth]{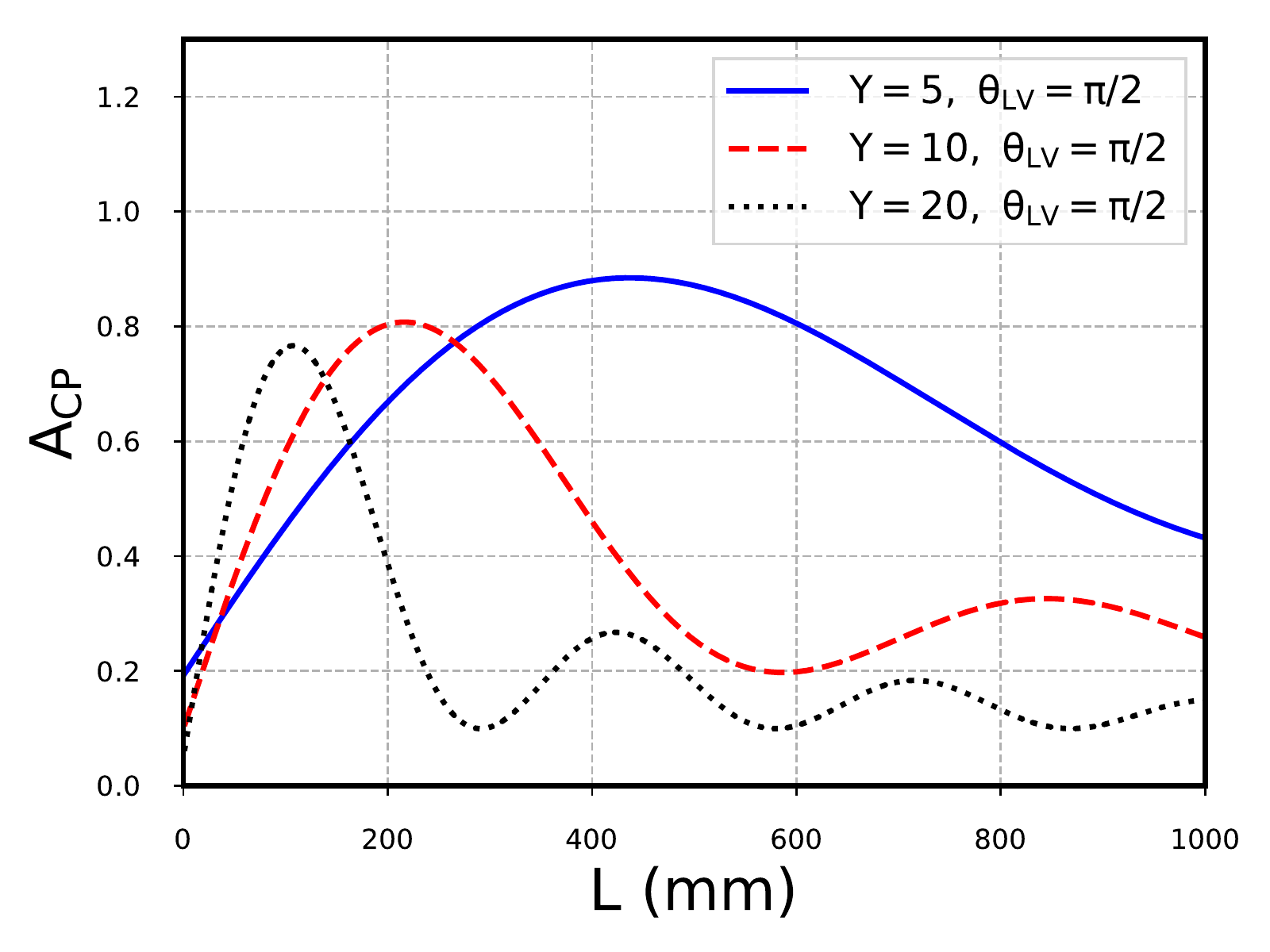}
\includegraphics[width=0.49\textwidth]{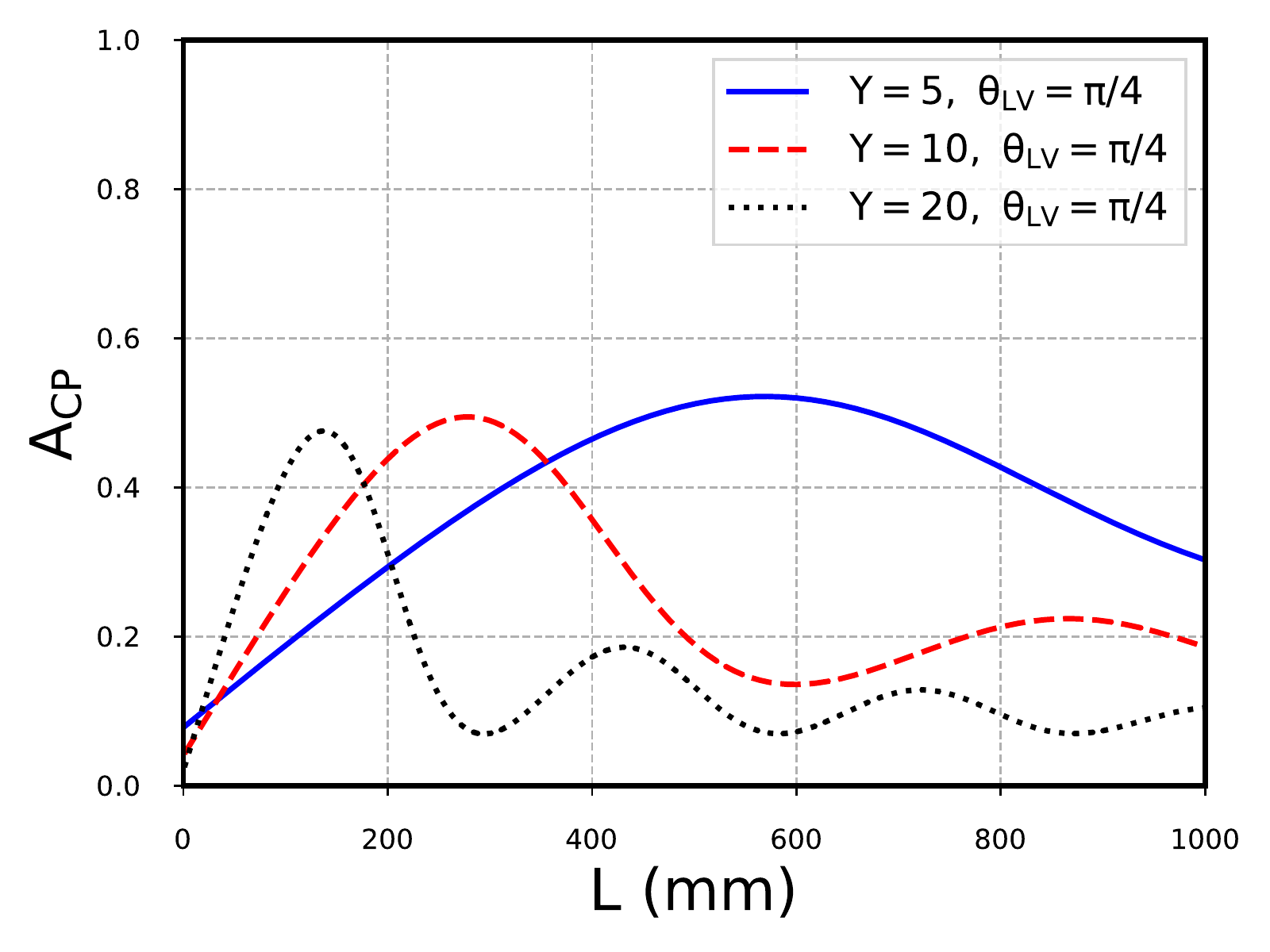}
\includegraphics[width=0.49\textwidth]{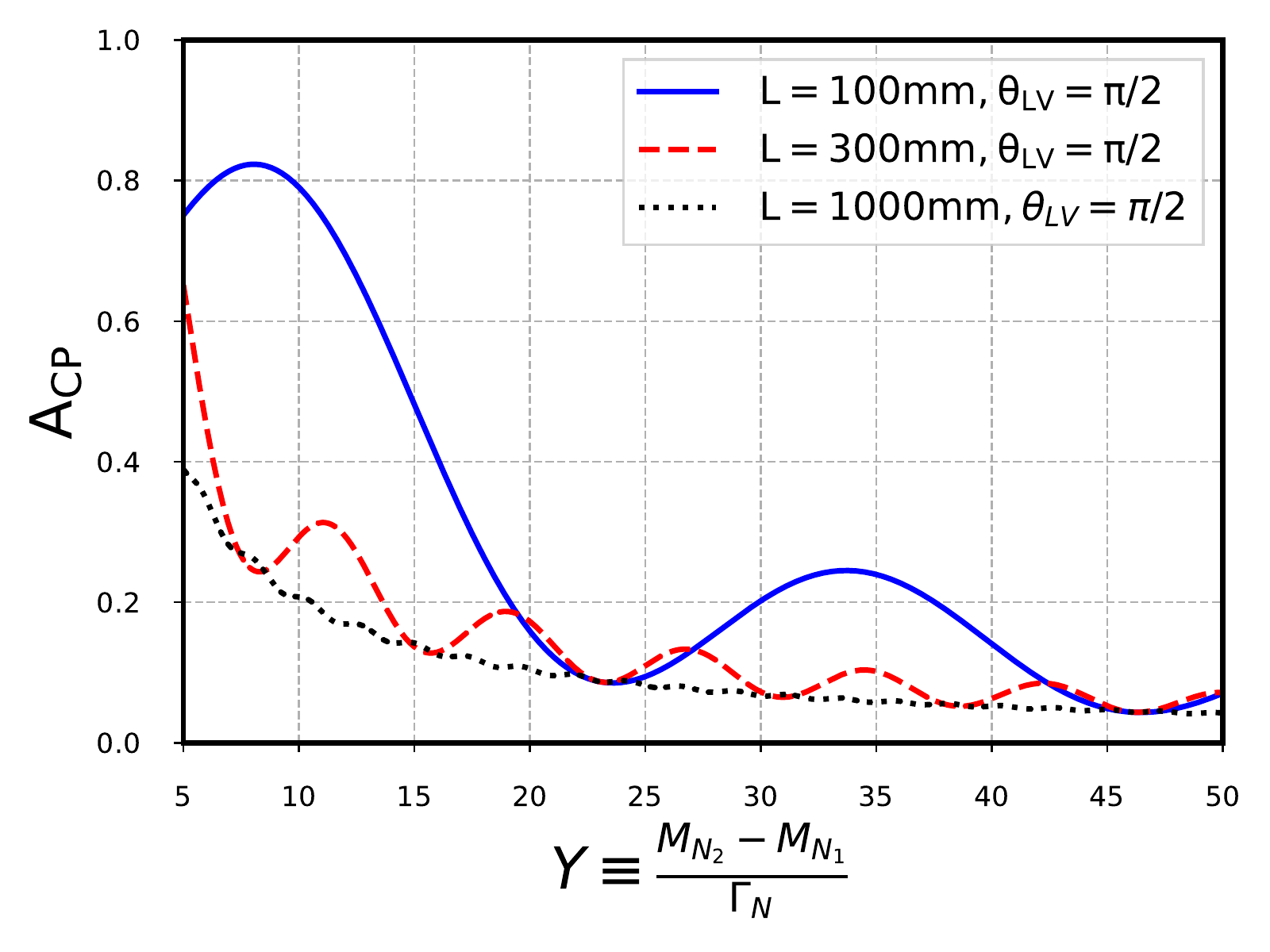}
\includegraphics[width=0.49\textwidth]{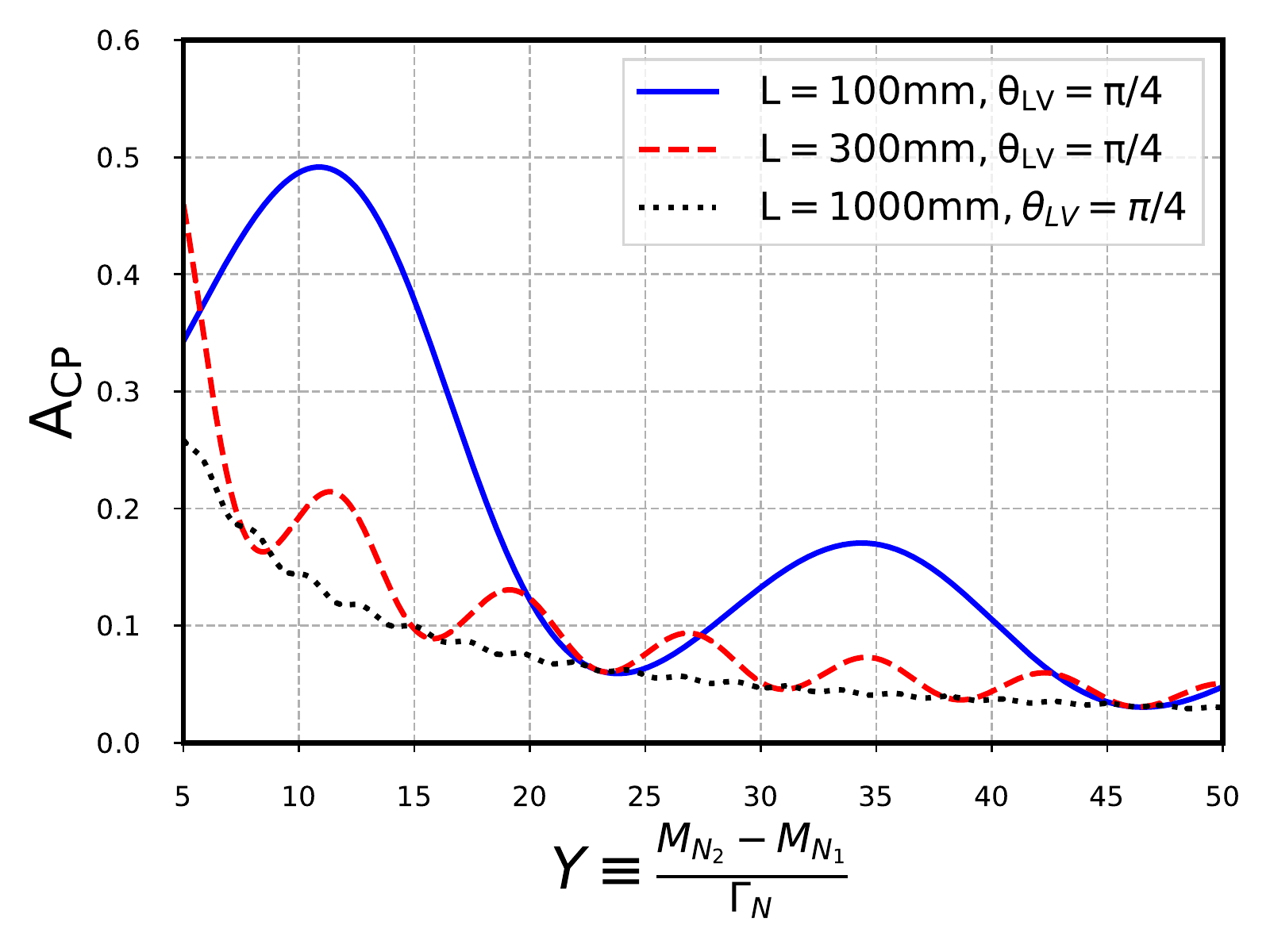}
\caption{CP asymmetry for a fixed value of $\gamma_N^{''} \beta_N^{''}$ ($= 664613$) (Eq.~\ref{ACP_exact}) for $M_N=2.0$ GeV and $|B_{\mu N}|^2=|B_{\tau N}|^2=10^{-5}$. Top panels show $\rm A_{CP}$ as a function of $L$, bottom panels as a function of $Y$. Left panels: $\theta_{LV}=\pi/2$. Right panels: $\theta_{LV}=\pi/4$. }
\label{fig:acp_exact}
\end{figure}

  On the other hand, some previous works (e.g. Ref.~\cite{Cvetic:2015naa,Cvetic:2016fbv}) have considered fixed values of $\gamma_N^{''} \beta_N^{''}$\footnote{Here we will consider $\gamma_N^{''} \beta_N^{''}= 0.664613$ which is the naive average value in the lab frame (when the weight function is constant).}; in this scenario the CP asymmetry (Eq.~\ref{acp}) can be approximated as follows

\begin{equation}
\label{ACP_exact}
A_{\rm CP} \approx \sin(\theta_{LV}) \frac{\Big[2-\exp \big( -L \Gamma_N/(\gamma_N^{''} \beta_N^{''}) \big)\Big( 1+\cos(\frac{2\pi L}{L_{\rm osc}})+\frac{1}{Y} \sin(\frac{2\pi L}{L_{\rm osc}})\Big)\Big]\frac{Y}{Y^2 + 1}}{\Big[ 1+\frac{2}{Y^2+1}\cos{\theta_{LV}}\Big]-\exp \big( -L \Gamma_N/(\gamma_N^{''} \beta_N^{''}) \big)\Big[ 1+\frac{\cos(\theta_{LV})}{Y^2 +1} \Big(1+\cos(\frac{2\pi L}{L_{\rm osc}})-Y \sin(\frac{2\pi L}{L_{\rm osc}}) \Big) \Big]}\ .
\end{equation}
Figure \ref{fig:acp_exact} shows the CP asymmetry in Eq.~\ref{ACP_exact} as a function of the maximal displaced vertex length $L$ (top) and $Y$ parameter (bottom)

\section{Discussion of the results and summary}
\label{sec:dis}

In this work we have studied the CP-violating effects in the rare $B$ meson decays mediated by two on-shell HNs. Unlike previous works, our calculations include both overlap (resonant) and oscillating effects. The variation of the values of the parameter $Y \equiv \Delta M_N/\Gamma_N$ shows that there exists a mass-difference regime in which the CP-violating effects can be noticeable. Our formulas are approximations which are good if $Y$ is not too small ($Y \gtrsim 5$), because we do not know (and do not include) the terms which are simultaneously overlap and oscillation effects. On the other hand, if $Y < 1$, i.e., the mass difference $\Delta M_N$ is smaller than the decay width $\Gamma_N$, the CP-violating effects are expected to be highly suppressed and ${\rm A_{CP}} \to 0$ as $Y\to 0$. We set the maximum value of the displaced vertex length (effective detector length) $L$ to $L=1000$ mm in order to obtain a realistic prediction of the number of events that can take place at Belle II experiment.  

While figures \ref{fig:compar} show that in both cases $Y=5,\ 10$ the oscillatory effects have a bigger contribution to the total effective branching ratio (full lines), we can see that both effects (oscillatory and overlap) contributions are of the same order of magnitude. In addition,  figure \ref{fig:compar} (top panel) we can see that the biggest difference from $B^+$ and $B^-$ effective branching ratios occurs between the $200$ and $400$ mm. Furthermore, the channel difference  changes with the CP violating phase $\theta_{LV}$, where the biggest CP violation appears at $\pi/2$ and the smallest occurs at $\pi/4$. For values of $\theta_{LV}=0,\pi$, there will be no difference between the channels. If the parameter $Y$ increases from 5 to 10 (figure \ref{fig:compar}, bottom panel) one can notice that now the biggest CP violation moves to the left and occurs between $50$ and $200$ mm, while the maximum occurs at $\theta_{LV}=\pi/2$. 

The effect produced by the parameter $Y$ can be read from figures \ref{fig:beff3} and \ref{fig:beff4}. Values of $Y>15$ shows little difference between the channels, this is well expected as for larger $Y$ the resonant and oscillating regimes will disappear when $\Delta M\propto Y\gg 1$. The maximum CP violation is strongly dependent on the length $L$, as seen from $L=300$ mm in figure \ref{fig:beff3} and $L=1000$ mm in figure \ref{fig:beff4}.

Figure \ref{fig:acp1} shows the asymmetry as a function of the length $L$. Although, the biggest value of the CP asymmetry appears for small values of the length ($L\sim 50 - 300$ mm), the branching ratios increase as $L\to 1000$ mm. Thus, biggest values of CP asymmetry are not enough to detect events. Therefore, the size of the branching ratios must also be taken into account in order to have a signal in the detector. Figure \ref{fig:acp2} shows the asymmetry as a function of $Y$. The biggest values of CP asymmetry appear for $Y=1-20$, and will disappear for $Y>50$. If we fix the value of $\gamma_N^{''} \beta_N^{''}$ (Figure \ref{fig:acp_exact}) we observe a clear oscillatory behaviour of $A_{CP}$. On the other hand, these effects are suppressed in figures \ref{fig:acp1} and \ref{fig:acp2} ($\gamma_N^{''} \beta_N^{''}$ variable), due to the several integrations (average) performed in the evaluation of $A_{CP}$.

Moreover, table \ref{tab:L} presents the expected number of events ${\rm N_e(B^{\pm})=\eta \times N_{B} \times Br_{eff}(B^{\pm})}$, considering that the number of $B$ mesons expected at Belle II is ${\rm N_{B} = 5 \times 10^{10}}$. In addition, while the track reconstruction efficiency is greater than 90\%, we will consider it to be $\eta = 30$\% in order to have a more conservative approach \cite{Kou:2018nap}.
\begin{table}[H]
  \begin{center}
    \begin{tabular}{|c|c|c|c|c|c|} 
    	\hline
      $L$ [mm] & Y & $\theta_{LV}$ & $\approx N_e(B^- \to D^0 \mu^- \mu^- \pi^+)$ & $\approx N_e(B^+ \to D^0 \mu^+ \mu^+ \pi^-)$ & $\Delta N_e\equiv N_e(B^-)-N_e(B^+)$\\
      \hline
      $\quad 300\quad $ & $\quad 5\quad$ & $\quad \pi/2\quad$ & $20$ & $3$ & $17$\\
      300 & 10 & $\pi/2$ & $17$ & $6$ & $11$\\
      300 & 5 & $\pi/4$ & $21$ & $7$ & $14$\\
      300 & 10 & $\pi/4$ & $16$ & $8$ & $8$\\
      1000 & 5 & $\pi/2$ & $35$ & $13$ & $22$\\
      1000 & 10 & $\pi/2$ & $30$ & $18$ & $12$\\
      1000 & 5 & $\pi/4$ & $33$ & $17$ & $16$\\
      1000 & 10 & $\pi/4$ & $29$ & $20$ & $9$\\
      \hline
    \end{tabular}
    \caption{Expected events at Belle II experiment. Here $|B_{\mu N}|^2=|B_{\tau N}|^2=10^{-5}$ and $M_N=2\ GeV$.}
\label{tab:L}
  \end{center}
\end{table}
In summary, in this work we studied the B-mesons decays $B^{\pm}  \to D^0 \mu^{\pm}_1 \mu^{\pm}_2 \pi^{\mp}$ at Belle II, considering a $1000$ mm effective detector length. We focused in a scenario with two almost-degenerate heavy neutrinos with masses around $M_N\sim 2$ GeV. The effective branching ratios were calculated by considering that the heavy neutrino total decay width is equal for both, as a consequence of the assumption that the heavy-light mixing coefficients satisfy $|B_{\ell N_1}| = |B_{\ell N_2}|$ ($\equiv |B_{\ell N}|^2$) for $\ell=\mu, \tau$. Further, we considered $|B_{\mu N}|^2 \sim |B_{\tau N}|^2 \sim 10^{-5} \gg |B_{e N}|^2$. The calculations were performed in a scenario that contains both the overlap (resonant) and oscillating CP-violating sources. We observed that the biggest difference of detectable events occurs for $Y=5$ and $\theta=\pi/2$ (Table~\ref{tab:L}). 

We established that for certain presently allowed regime of values of $|B_{\mu N}|^2$, $Y (\equiv \Delta M_N/\Gamma_N)$ and $\theta_{LV}$, and with $M_N \approx 2$ GeV, the aforementioned effects can be observed at Belle II.

\section{Acknowledgments}
This work was supported in part by FONDECYT (Chile) Grants No.~1180344 (G.C.) and No.~3180032 (J.Z.S.).
The work of C.S.K. was supported by the National Research Foundation of Korea (NRF) 
grant funded by Korea government of the Ministry of Education, Science and Technology (MEST) 
(No. 2018R1A4A1025334).
\newpage
 \appendix

 \section{Decay width $\Gamma(B \to D \ell_1 N)$}
\label{sec:appA}
\begin{figure}[H]
\centering\includegraphics[width=90mm]{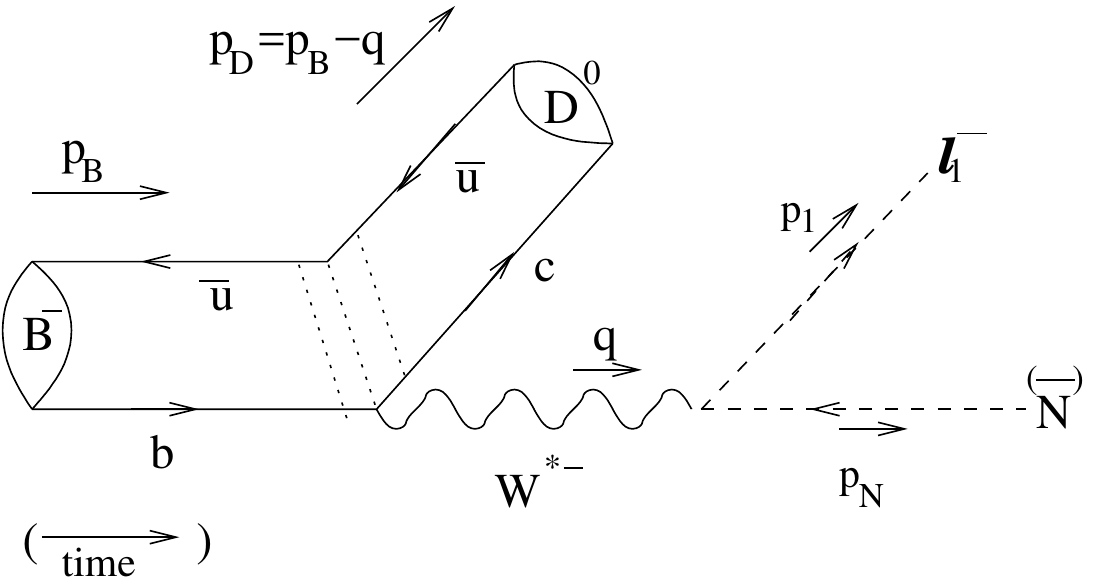}
\caption{Schematical representation of the decay $B^- \to D^0 \ell_1^- {\bar N}$ \cite{Cvetic:2016fbv}. }
\label{FigBDW}
\end{figure}
The differential decay width of the process $B \to D \ell_1 N$ (see Fig.~\ref{FigBDW}) was obtained in Ref.~\cite{Cvetic:2017vwl} and has the following form:\footnote{In Ref.~\cite{Cvetic:2017vwl} there is a typo in Eq.~(11) for this differential decay width, the expression given there must be multiplied by 4. The correct formula was used in the calculations there, though, which reproduces the decay width $\Gamma(B \to D \ell_1 N)$ calculated earlier in Ref.~\cite{Cvetic:2016fbv}.}
\bes
\label{dGBDlN}
\bea
\frac{d \Gamma(B \to D \ell_1 N)}{d q^2 d \Omega_{{\hat q}'} d \Omega_{{\hat p}_1}} &=& |B_{\ell_1 N}|^2 \frac{d \bG(B \to D \ell_1 N)}{d q^2 d \Omega_{{\hat q}'} d \Omega_{{\hat p}_1}}\ ,
\\
& = & \frac{|B_{\ell_1 N}|^2 |V_{c b}|^2 G_F^2}{ M_B (4 \pi)^5} |{\widetilde {\cal T}}|^2  \lambda^{1/2} \left( 1, \frac{M_D^2}{M_B^2}, \frac{q^2}{M_B^2} \right)  \lambda^{1/2} \left( 1, \frac{M_1^2}{q^2}, \frac{M_N^2}{q^2} \right).
\eea \ees
We denote the $W^{*}$-rest frame (i.e., $\ell_1 N$-rest frame) as $\Sigma$, and the $B$-rest frame as $\Sigma'$ (where the corresponding momenta have a prime).
In Eqs.~(\ref{dGBDlN}), $q^2$ is the squared four-momentum of the $W^{*}$ boson, ${\hat q}'$ is the unitary direction vector of ${\vec q}'$ in the $B$-rest frame $\Sigma'$, ${\hat p}_1$ is the unitary direction of ${\vec p}_1$ of $\ell_1$ in the $W^*$-rest ($\ell_1 N$-rest) frame $\Sigma$, in fact $d\Omega_{\hat p_1}=d\phi_1 d(\cos\theta_1)$.
The expression $|{\widetilde {\cal T}}|^2$ stands for the squared decay amplitude and is given by
\begin{align}
\nonumber
|{\widetilde {\cal T}}|^2 =&
\frac{1}{q^2} F_1(q^2) (F_0(q^2)-F_1(q^2)) \left(M_B^2-M_D^2\right) \\
\nonumber
   &{\bigg [}  M_1^2 \left(-4 (\cos \theta_1 |{\vec p}_D| |{\vec p}_N|+p_D^0 p_1^0)+2 M_B^2-2 M_D^2+2 M_N^2 -q^2\right)
\nonumber\\
     & +M_N^2 \left(4 (\cos \theta_1 |{\vec p}_D| |{\vec p}_N|+p_D^0 p_1^0)-M_N^2+q^2\right)-M_1^4 {\bigg ]}
\nonumber\\
& -\frac{1}{2} F_1(q^2)^2 {\bigg [}M_1^2
   \left( 8 (\cos \theta_1 |{\vec p}_D|
   |{\vec p}_N|+p_D^0 p_1^0)-4 M_B^2-2 M_N^2+3
   q^2\right)
   \nonumber\\
&   -8 M_B^2 (\cos \theta_1 |{\vec p}_D| |{\vec p}_N|+
   p_D^0 p_1^0)+M_D^2 \left(8 (\cos \theta_1 |{\vec p}_D|
|{\vec p}_N|+p_D^0 p_1^0)-4 M_N^2+4 q^2\right)
 \nonumber\\
& -8 M_N^2 (\cos \theta_1 |{\vec p}_D| |{\vec p}_N|+p_D^0 p_1^0)
+8q^2 (\cos \theta_1 |{\vec p}_D| |{\vec p}_N|+p_D^0 p_1^0)
 \nonumber\\
& +16 (\cos \theta_1 |{\vec p}_D| |{\vec p}_N|+p_D^0
   p_1^0)^2+M_1^4+M_N^4-M_N^2
   q^2 {\bigg ]}
\nonumber\\
& +\frac{1}{2 (q^2)^2}
(F_0(q^2)-F_1(q^2))^2 \left(M_B^2-M_D^2\right)^2 \left[-M_1^4+M_1^2 \left(2
     M_N^2+q^2\right)-M_N^4+M_N^2 q^2\right] \ ,
\label{sqamp}
\end{align}
where
\bes
\label{vecpo0}
\bea
|{\vec p}_N| = |{\vec p}_1| & = & \frac{1}{2} \sqrt{q^2}  \; \lambda^{1/2} \left( 1, \frac{M_1^2}{q^2}, \frac{M_N^2}{q^2} \right)\ ,
\label{vecpN}
\\
|{\vec p}_D| & = & \frac{M_B^2}{2 \sqrt{q^2}} \; \lambda^{1/2} \left( 1, \frac{M_D^2}{M_B^2}, \frac{q^2}{M_B^2} \right) = \frac{M_B |{\vec {q'}}|}{\sqrt{q^2}}\ ,
\label{vecpD}
\\
p_1^0 & = & \frac{1}{2 \sqrt{q^2}} (q^2 - M_N^2 + M_1^2)\ ,
\label{p10}
\\
p_D^0 & = & \frac{1}{2 \sqrt{q^2}} (M_B^2 - M_D^2 - q^2)\ .
\label{pD0}
\eea
\ees
These momenta are all in the $W^{*}$-rest frame ($\Sigma$); $\theta_1$ is the angle between ${\vec p}_1$ and ${\hat z} \equiv {\hat q}'$ (the direction of $W^{*}$ in the $B$-rest frame).

The expression (\ref{sqamp}) is defined in terms of two form factors, $F_1$ and $F_0$. The form factor $F_1(q^2)$ is presented in \cite{Caprini:1997mu} and is expressed in terms of $w(q^2)$ and $z(w)$
\bes
\label{wz}
\bea
w(q^2) & = & \frac{(M_B^2 + M_D^2 - q^2)}{2 M_B M_D} \ ,
\label{w}
\\
z(w) & = & \frac{\sqrt{w+1} - \sqrt{2}}{\sqrt{w+1} + \sqrt{2}} \ .
\label{z}
\eea
\ees
Therefore, from Ref.~\cite{Caprini:1997mu}, $F_1(q^2)$ can be expressed as
\be
F_1(q^2) = F_1(w=1) \left( 1 - 8 \rho^2 z(w) + (51 \rho^2 - 10) z(w)^2 - (252 \rho^2 - 84) z(w)^3 \right) \ .
\label{CLNF1}
\ee
In the last equation the free parameters $\rho^2$ and $F_1(w=1)$ have been determined by the Belle Collaboration \cite{Glattauer:2015teq}
\bes
\label{rho2F1max}
\bea
\rho^2 &= & 1.09 \pm 0.05 \ ,
\label{rho2}
\\
|V_{cb}| F_1(w=1) &=& (48.14 \pm 1.56) \times 10^{-3} \ .
\label{F1max}
\eea
\ees
The form factor $F_0(q^2)$ is given as \cite{Caprini:1997mu}\footnote{In Ref.~\cite{Cvetic:2016fbv}, $f_0(w)$ was transcribed there in Eq.~(11b) with a typo [$+ {\rho}_0^2 (w - 1)$ instead of  $-{\rho}_0^2 (w - 1)$], but the correct expression (\ref{F0b}) was used in the calculations there.}
\bes
\label{F0}
\bea
F_0(q^2) & = & \frac{(M_B+M_D)}{2 \sqrt{M_B M_D}}
\left[ 1 - \frac{q^2}{(M_B+M_D)^2} \right] f_0(w(q^2)) \ ,
\label{F0a}
\\
f_0(w) & \approx & f_0(w=1) \left[ 1 - {\rho}_0^2 (w - 1) + (0.72 \rho_0^2 - 0.09) (w - 1)^2 \right] \ ,
\label{F0b}
\eea
\ees
where $f_0(w=1) \approx 1.02$ and $\rho_0^2 \approx 1.102$.

The decay width for $B \to D \ell_1 N$ decays is
\bea
\Gamma(B \to D \ell_1 N) & = &
    |B_{\ell_1 N}|^2 \int d q^2 \int d \Omega_{{\hat q}'} \int d \Omega_{{\hat p}_1}
    \frac{d \bG(B \to D \ell_1 N)}{ d q^2 d \Omega_{{\hat q}'}  d \Omega_{{\hat p}_1}} \ ,
\label{GBDlN}
\eea
For the effective decay width, which takes into account only those decays in which the exchanged on-shell $N$ decays within the detector, we refer to Appendix \ref{sec:appC}.

\section{Decay width for $N \to \ell^{\pm} \pi^{\mp}$}
\label{sec:appB}

\begin{figure}[H]
\centering
\includegraphics[scale = 0.85]{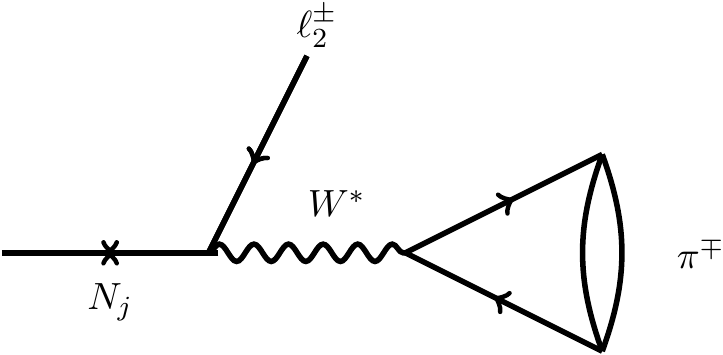}
\caption{Feynmann diagram for the decay process $N \to \ell^{\pm} \pi^{\mp}$.}
\label{fig:Nellpi}
\end{figure}

The decay width $\Gamma(N \to \ell^{\pm} \pi^{\mp})$ (see Fig.~\ref{fig:Nellpi}) is proportional to the heavy-light mixing factor $|U_{\ell_2 N}|^2$
\be
\Gamma(N \to \ell^{\pm} \pi^{\mp}) = |B_{\ell_2 N}|^2 \bG(N \to \ell^{\pm} \pi^{\mp}) \ .
\label{GNlPi}
\ee
Here, the canonical decay width $\bG$ is
\be
\bG(N \to \ell^{\pm} \pi^{\mp}) =
\frac{1}{16 \pi} |V_{u d}|^2 G_F^2 f_{\pi}^2 M_N^3 \lambda^{1/2}(1, x_{\pi}, x_{\ell})
\left[ 1 - x_{\pi} - 2 x_{\ell} - x_{\ell}  (x_{\pi}-x_{\ell}) \right] \ ,
\label{bGNlPi}
\ee
where $f_{\pi}$ ($\approx 0.1304$ GeV) is the decay constant of pion,
and the other factors are
\be
x_{\pi} = \frac{M_{\pi}^2}{M_N^2} \ , \qquad x_{\ell}=\frac{M_{\ell}^2}{M_N^2}\ .
\label{xPixell}
\ee

These results can be combined with the result (\ref{GBDlN}) to obtain the decay width for the decay $B^{\pm} \to D^0 \ell_1^{\pm} N \to \ell_1^{\pm} \ell_2^{\pm} \pi^{\mp}$
\bea
  \Gamma(B^{\pm} \to D^0 \ell_1^{\pm} N \to \ell_1^{\pm} \ell_2^{\pm} \pi^{\mp}) &=&
  \Gamma(B^{\pm} \to D^0 \ell_1^{\pm} N) \frac{\Gamma(N \to \ell_2^{\pm} \pi^{\mp})}{\Gamma_N}\ ,
  \nonumber\\ &= & 
  |B_{\ell_1 N}|^2 |B_{\ell_2 N}|^2 \frac{ \bG(N \to \ell_2 \pi) }{\Gamma_N}
  \int d q^2 \int d \Omega_{{\hat q}'} \int d \Omega_{{\hat p}_1}
 \frac{d \bG(B \to D \ell_1 N)}{ d q^2 d \Omega_{{\hat q}'}  d \Omega_{{\hat p}_1}}\ ,
 \label{GBDllpi}
 \eea 
where the expressions (\ref{dGBDlN}) and (\ref{GBDlN}) are used for the first factor, and (\ref{GNlPi}) and (\ref{bGNlPi}) for the second factor of the integrand. For $\Gamma_N$ we refer to Eq.~(\ref{DNwidth}).

\section{Lorentz factors of on-shell $N$ in laboratory frame}
\label{sec:appC}

In this Appendix we follow the presentation given in Ref.~\cite{Cvetic:2017vwl}. The expression (\ref{GBDllpi}) refers to the decay width for all the decays of the type $B^{\pm} \to D^0 \ell_1^{\pm} N \to \ell_1^{\pm} \ell_2^{\pm} \pi^{\mp}$, including those where the on-shell $N$ decays outside the detector. However, if we realistically consider that only those decays are detected in which the on-shell $N$ decays within the detector (of length $L$), we need to multiply the integrand in Eq.~(\ref{GBDllpi}) with the probability $P_N$ of decaying of the produced on-shell $N$ within the length $L$.
\be
P_N = 1 - \exp \left[ - \frac{L}{\tau_N \gamma_N^{''} \beta_N^{''}} \right]
= 1 - \exp \left[ - \frac{L \Gamma_N}{\gamma_N^{''} \beta_N^{''}} \right]\ ,
\label{PN}
\ee 
where $\tau_N = 1/\Gamma_N$ is the lifetime of $N$ in its rest frame. The velocity $\beta_N^{''}$ and the Lorentz factor $\gamma_N^{''} = 1/\sqrt{1-(\beta_N^{''})^2}$ are those of the $N$ neutrino in the lab frame $\Sigma''$.\footnote{We use the same conventions as in Appendix \ref{sec:appA}: the $W^{*}$-rest frame ($\ell_1 N$-rest frame) is $\Sigma$, and the $B$-rest frame is $\Sigma'$. The lab frame is denoted as $\Sigma''$ (and the corresponding momenta have double prime). Note, however, that the distance between the two vertices of the on-shell $N$ in the lab frame is denoted for simplicity as $L$ (and not: $L^{''}$).}

At Belle II, the kinetic energy of the produced $\Upsilon(4S)$ is $0.421$ GeV, and this implies that its Lorentz factor in the lab frame $\Sigma''$ is $\gamma_{\Upsilon}^{''}=1.0398$ and $\beta_{\Upsilon}^{''}=0.274$. When $\Upsilon(4S)$ produces a pair of $B$ mesons, the kinetic energy of $B$ mesons in the $\Upsilon$-rest frame is $0.010$ GeV, which is negligible. Thus the velocity of the $B$ mesons in the lab frame $\Sigma''$ is equal to the velocity of $\Upsilon(4S)$
\be
\beta_B^{''}=0.274\ , \qquad \gamma_B^{''} = 1.0398\	 .
\label{LorB} \ee 
Then, the factor $\gamma_N^{''} \beta_N^{''}$ appearing in the probability (\ref{PN}) can be calculated by calculating the energy $E_N^{''}$ of the $N$ neutrino in the lab frame (see below)
\be
\gamma_N^{''} \beta_N^{''} = \sqrt{ \left(E''_N(q^2;{\hat q}',{\hat p}_{1})/M_N \right)^2 - 1 }\ ,
\label{bNgNpp}
\ee 
and this leads to the effective decay width for the considered process
  \bea
{\Gamma}_{\rm eff}(B \to D \ell_1 N \to D \ell_1 \ell_2 \pi) & = &
|B_{\ell_1 N}|^2 |B_{\ell_2 N}|^2  \frac{ \bG(N \to \ell_2 \pi) }{\Gamma_N}
\int d q^2 \int d \Omega_{{\hat q}'} \int d \Omega_{{\hat p}_1}
    \frac{d \bG(B \to D \ell_1 N)}{ d q^2 d \Omega_{{\hat q}'}  d \Omega_{{\hat p}_1}} 
\nonumber\\
&& \times \left\{ 1 - \exp \left[- \frac{L \Gamma_N}{\sqrt{ \left(E''_N(q^2;{\hat q}',{\hat p}_{1})/M_N \right)^2 - 1 }} \right] \right\}\ ,
\label{Geff}
\eea 
which is as the expression (\ref{GBDllpi}) but with inclusion of the $N$ decay probability within the effective detector length $L$.\footnote{The effective detector length here is considered to be independent of the position of the $N$-production vertex and independent of the direction in which the produced $N$ travels.}
The energy $E''_N$ of the produced heavy neutrino $N$ in the lab frame and is given by (cf.~App.~B of Ref.~\cite{Cvetic:2017vwl})
\begin{align}
\label{ENppfin}
E''_N(q^2; \theta_q; \theta_1, \phi_1)\ =\ & \gamma_B \Bigg(
\gamma_W(q^2) \Big( E_N(q^2) - \beta_W(q^2) |{\vec p}_N(q^2)| \cos \theta_1 \Big) \\
\nonumber
& + \beta_B \Big[ \gamma_W(q^2) \Big( - |{\vec p}_N(q^2)| \cos \theta_1 + \beta_W(q^2) E_N(q^2) \Big) \cos \theta_q \\ \nonumber
& - |{\vec p}_N(q^2)|\sin \theta_1 \cos \phi_1 \sin \theta_q \Big] \Bigg)\ .
\end{align}
The factors, as a function of the squared invariant mass of $W^{*}$, $q^2$ (see Fig.~\ref{FigBDW}), are
\be
\nonumber
E_N  =  \frac{1}{2 \sqrt{q^2}} (q^2 + M_N^2 - M_1^2)\ ,
\label{EN}
\ee

\be
\nonumber
|{\vec p}_N| = |{\vec p}_1| = \frac{1}{2} \sqrt{q^2} \lambda^{1/2} \left( 1, \frac{M_1^2}{q^2}, \frac{M_N^2}{q^2} \right)\ ,
\label{vecpNp1}
\ee

\be
\nonumber
\gamma_W(q^2) = \left( 1 + \frac{|{\vec {q'}}|^2}{q^2} \right)^{1/2}\ , \qquad
\beta_W(q^2) = \left( \frac{q^2}{|{\vec {q'}}|^2} + 1 \right)^{-1/2}\ ,
\label{gWbW}
\ee
and
\be
\nonumber
|{\vec {q'}}| = \frac{1}{2} M_B \lambda^{1/2} \left( 1, \frac{ M_{\Dst}^2}{M_B^2}, \frac{q^2}{M_B^2} \right)\ ,
\label{vecq}
\ee
see also Eqs.~(\ref{vecpo0}). In Eq.~(\ref{ENppfin})
the angles $\theta_q$, $\theta_1$ and $\phi_1$ are in the following ranges:
\begin{align*}
0 \leq  \theta_q \leq \pi\ , & \\
0 \leq  \theta_1 \leq \pi\ , & \\
0 \leq  \phi_1 < 2 \pi & \ .  
\end{align*}
For a more detailed explanation of the aforemention expressions we refer to Ref.~\cite{Cvetic:2017vwl}.

\section{Effective width of the LNV $B$ decay channel with overlap and oscillation effects}
\label{sec:appDW}

Here we will explain how the expression (\ref{DWeff}) is obtained. We work in the case when the Lorentz factors in the lab frame $\beta_N^{''}$ and $\gamma_N^{''} \equiv 1/\sqrt{1-(\beta_N^{''})2}$ are considered to be fixed. In addition, we use the assumption made throughtout this work that the heavy-light mixing elements satisfy $|B_{\ell N_1}| = |B_{\ell N_2}|$ ($\equiv |B_{\ell N}|$), where $\ell=\mu, e, \tau$. When no oscillation is assumed [i.e., only the overlap (resonant) effects included], the effective decay width for the considered LNV decay channnel is \cite{Cvetic:2014nla} [cf.~also \cite{Cvetic:2015ura} Eq.~(13) there]
\bea
\Gamma_{\rm eff}(B^{\pm} \to D^0 \ell^{\pm}_1 \ell^{\pm}_2 \pi^{\mp})_{\rm res} &=&  \Gamma(B^{\pm} \to D^0 \ell^{\pm}_1 N) \ \frac{\Gamma(N \to \ell^{\pm}_2 \pi^{\mp})}{\Gamma_N} 2 |B_{\ell_1 N}|^2  |B_{\ell_2 N}|^2
\nonumber \\ 
& & \times \left[ 1- \exp \left( - L \Gamma_N/( \gamma_N^{''} \beta_N^{''} ) \right) \right] \left[ 1 + \delta(Y) \cos(\theta_{LV}) \mp \frac{\eta(Y)}{Y} \sin(\theta_{LV}) \right].
\label{Geffres} \eea
We recall that $L$ here is the length of flight of the on-shell $N_j$ in the detector before it decays (within the detector), and the parameter $Y$ and the $N_1$-$N_2$ overlap functions $\delta(Y)$ and $\eta(Y)$ are given in Eqs.~(\ref{Ydef}) and (\ref{etadel}. The differential decay rate $d \Gamma_{\rm eff}/d L$ for this decay width is then
\bea
\left(\frac{d \Gamma_{\rm eff}}{d L} \right)_{\rm res} & = & \frac{1}{\gamma_N^{''} \beta_N^{''}} \exp \left( - \frac{L \Gamma_N}{\gamma_N^{''} \beta_N^{''}}  \right)  \Gamma(B^{\pm} \to D^0 \ell^{\pm}_1 N) \Gamma(N \to \ell^{\pm}_2 \pi^{\mp}) 2 |B_{\ell_1 N}|^2  |B_{\ell_2 N}|^2
\nonumber \\ && \times
\left[ 1 + \delta(Y) \cos(\theta_{LV}) \mp \frac{\eta(Y)}{Y} \sin(\theta_{LV}) \right]\ .
\label{dGeffres} \eea
On the other hand, when $Y \gg 1$ and thus the overlap contributions $\sim \delta(Y)$ and $\sim \eta(Y)/Y$ can be neglected, we obtained in Ref.~\cite{Cvetic:2015ura} the corresponding differential decay width with $N_1$-$N_2$ oscillation effects included\footnote{In \cite{Cvetic:2015ura} we wrote this expression in the approximation of small $N_j$-decay probability $P_N(L) \equiv [1 - \exp(-L \Gamma_N/(\gamma_N^{''} \beta_N^{''}))]$, namely $P_N(L) \approx L \Gamma_N/(\gamma_N^{''} \beta_N^{''})$. In Refs.~\cite{Cvetic:2018elt,Cvetic:2019rms} and here we wrote this expression without this approximation, which gives us an additional factor $\exp(-L \Gamma_N/(\gamma_N^{''} \beta_N^{''}))$ in $d P_N(L)/dL$ and in $(d \Gamma_{\rm eff}/d L)_{\rm osc}$.}
\bea
\left(\frac{d \Gamma_{\rm eff}}{d L} \right)_{\rm osc} & = & \frac{1}{\gamma_N^{''} \beta_N^{''}} \exp \left( - \frac{L \Gamma_N}{\gamma_N^{''} \beta_N^{''}}  \right)  \Gamma(B^{\pm} \to D^0 \ell^{\pm}_1 N) \Gamma(N \to \ell^{\pm}_2 \pi^{\mp}) 2 |B_{\ell_1 N}|^2  |B_{\ell_2 N}|^2
\nonumber \\ && \times
\left[ 1 + \cos \left( 2 \pi \frac{L}{L_{\rm osc}} \pm \theta_{LV} \right) \right]\ ,
\label{dGeffosc} \eea
where $L_{\rm osc}$ is the HN oscillation length
\be
L_{\rm osc} = \frac{2\pi \gamma_N^{''} \beta_N^{''}}{\Delta M_N} \; \Rightarrow
2 \pi \frac{L}{L_{\rm osc}} = Y \frac{\Gamma_N}{\gamma_N^{''} \beta_N^{''}} L\ .
\label{Losc} \ee
If we now combine the overlap (resonant) contributions contained in the expression (\ref{dGeffres}) with the oscillation contributions contained in the expression (\ref{dGeffosc}), we obtain
\bea
\left(\frac{d \Gamma_{\rm eff}}{d L} \right) & = & 
\frac{1}{\gamma_N^{''} \beta_N^{''}} \exp \left( - \frac{L \Gamma_N}{\gamma_N^{''} \beta_N^{''}}  \right)  \Gamma(B^{\pm} \to D^0 \ell^{\pm}_1 N) \Gamma(N \to \ell^{\pm}_2 \pi^{\mp}) 2 |B_{\ell_1 N}|^2  |B_{\ell_2 N}|^2
\nonumber\\
&& \times {\Big \{}  1 + \left[ \delta(Y) \cos(\theta_{LV}) \mp \frac{\eta(Y)}{Y} \sin(\theta_{LV}) \right] +  \cos \left( 2 \pi \frac{L}{L_{\rm osc}} \pm \theta_{LV} \right) {\Big \}}\ .
\label{dGeff} \eea
The expression (\ref{dGeffosc}) was obtained in Ref.~\cite{Cvetic:2015ura} from the expression (\ref{dGeffres}) under the assumption that the overlap contributions ($\sim \delta(Y), \eta(Y)/Y$) there were negligible, i.e., that $Y \gg 1$.
Combination of these two expressions into the expression (\ref{dGeff}) thus involves an approximation of neglecting oscillation terms which involve overlap effects, i.e., terms of the type $\sim (\eta(Y)/Y) \cos( 2 \pi L/L_{\rm osc}  \pm \theta_{LV} )$ or similar (we do not know these terms).\footnote{One may be worried that this hierarchical view may not be adequate, because Eq.~(\ref{dGeff}) for $d \Gamma_{\rm eff}/dL$ may suggest that the overlap effects (at $Y \gg 1$) are smaller than the oscillation effects. Nonetheless, the two types of effects are mutually comparable in the integrated width $\Gamma_{\rm eff}$ (cf.~also the last paragraph in this Appendix).} This approximation is also reflected in the fact that the expression (\ref{dGeff}) is negative for some flight lengths $L$, which should not happen. However, if $Y$ is significantly larger than one (say, $Y \gtrsim 5$), these negative contributions are small in absolute value and appear only in very short intervals of $L$, and consequently the expression (\ref{dGeff}) can be regarded as a reasonably good approximation containing simultaneously both the overlap (resonant) and oscillation contributions, especially when it is integrated over $L$.

Integration of the partial decay width (\ref{dGeff}) from $L=0$ to $L$ then gives us the expression (\ref{DWeff}) in Sec.~\ref{s2} [expresssion (\ref{DWeffgen}) in Sec.~(\ref{sec:simu})], where we can clearly see that the overlap contribution and the oscillation contribution to $\Gamma_{\rm eff}(B \to D \ell_1 \ell_2 \pi)$ are mutually comparable, in view of the relations (\ref{etadel}).


\bibliographystyle{apsrev4-1}
\bibliography{biblio.bib}

\end{document}